\documentclass[12pt]{article}
\usepackage{epsf}
\usepackage{cite,,epsfig}
\usepackage{graphicx}

\newcommand{\beq}{\begin{equation}}
\newcommand{\eeq}{\end{equation}}
\newcommand{\bea}{\begin{eqnarray}}
\newcommand{\eea}{\end{eqnarray}}

\newcommand{\sla}[1]%
        {\kern .25em\raise.18ex\hbox{$/$}\kern-.75em #1}
\newcommand{\mybar}[1]%
        {\kern 0.8pt\overline{\kern -0.8pt#1\kern -0.8pt}\kern 0.8pt}
\newcommand{\gsim}{\, \raisebox{-0.8ex}{$\stackrel{\textstyle >}{\sim}$ }}

\newcommand{\BR}{\mathcal{B}}

\newcommand{\Btaun}{{B \to \tau \nu}}

\newcommand{\no}{\nonumber}

\begin{document}

\begin{center}

{\Large\bf Anatomy and Phenomenology of the Lepton Flavor Universality in SUSY Theories}
  \\ [25 pt]
{\sc {\sc A.~Masiero${}^{a}$, P.~Paradisi${}^{b}$, and R.~Petronzio${}^{c}$ }
  \\ [25 pt]

{\sl ${}^a$ Dip. di Fisica `G. Galilei', Univ. di Padova and 
INFN, Sezione di Padova, Via Marzolo 8, I-35131, Padua, Italy} \\ [5 pt] 

{\sl ${}^b$Physik-Department, Technische Universit\"at M\"unchen,
D-85748 Garching, Germany} \\ [5 pt]

{\sl ${}^c$Dip. di Fisica, Universit\`a di Roma
``Tor Vergata'' and INFN, Sezione di Roma, ``Tor Vergata'',
Via della Ricerca Scientifica 1, I-00133, Rome, Italy} \\ [25 pt]  }

{\bf Abstract} \\
\end{center}

\noindent

High precision electroweak tests, such as deviations from the Standard Model expectations
of the Lepton Flavor Universality breaking in $K\rightarrow \ell\nu_{\ell}$
(with $l= e$ or $\mu$), represent a powerful tool to test the Standard Model and, hence,
to constrain or obtain indirect hints of New Physics beyond it. We explore such a possibility
within Supersymmetric theories. Interestingly enough, a process that in itself does not need 
lepton flavor violation to occur, i.e. the violation of $\mu-e$ non-universality in $K_{\ell2}$,
proves to be quite effective in constraining not only relevant regions of SUSY models where
lepton flavor is conserved, but even those where specific lepton flavor violating contributions
arise. Indeed, a comparison with analogous bounds coming from $\tau$ lepton flavor violating
decays shows the relevance of the measurement of $R_{K}$ to probe Lepton Flavor Violation in SUSY.
We outline the role and the interplay of the direct New Physics searches at the LHC with the
indirect searches performed by LFU tests.

\vskip 1.5 cm

\section{Introduction}

The study of Lepton Flavor Universality (LFU) represents a powerful tool to test
the Standard Model (SM) and, hence,
to constrain or obtain indirect hints of new physics beyond it. Kaon and pion
physics are obvious grounds where to perform such tests, for instance in the
$\pi\rightarrow\ell\nu_{\ell}$ and $K\rightarrow\ell\nu_{\ell}$ decays, where
$\ell= e$ or $\mu$.
In particular, defining $(R_P^{e/\mu})_{SM}=\Gamma(P\to e\nu_e)_{SM}/\Gamma(P\to\mu\nu_e)_{SM}$
and $(R_P^{e/\mu})_{exp.}=\Gamma(P\to e\nu)_{exp.}/\Gamma(P\to \mu\nu)_{exp.}$,
the difference of the ratio
\beq
R_P^{e/\mu}  =
\frac{(R_P^{e/\mu})_{exp.}}{(R_P^{e/\mu})_{SM}}
= 1 + \Delta r_P^{e/\mu}
\label{RPemu}
\eeq
from unit signals the presence of LFU violating New Physics (NP).
Given that $(R_P^{e/\mu})_{SM}$ is accurately predicted,
 both for $P=\pi$ (0.02\% accuracy \cite{Marciano}) and $P=K$ (0.04\% 
accuracy \cite{Marciano}), it turns out that the determination of
$(R_P^{e/\mu})$ constitutes a major precision test of the SM.

These precision tests are equally interesting and fully complementary to the
flavor-conserving electroweak precision tests and to the FCNC tests performed
in hadronic and leptonic physics (rare kaon, charm and B physics, lepton 
Flavor Violation (LFV)): the smallness of NP effects is more than compensated in
terms of NP sensitivity by the excellent experimental resolution and the good 
theoretical control. The limiting factor in the determination $R_K^{e/\mu}$ is 
the $K\to e\nu$ rate, whose experimental knowledge has been quite poor so far.

The current world average $(R_K^{e/\mu})_{exp.} = (2.45 \pm 0.11)\times 10^{-5}$
~\cite{PDG} will be soon improved thanks to a series of preliminary results
by NA48/2 and KLOE (see Fig.~\ref{fig:susylimit}).
The two results by NA48/2, being based on different data sets (2003~\cite{flavianet}
and 2004~\cite{flavianet}, respectively) with different running conditions, should
be regarded as completely independent.
Combining these new results with the PDG value yields~\cite{flavianet}
\beq
(R_K^{e/\mu})_{exp.} = ( 2.457 \pm 0.032 ) \times 10^{-5}~.
\label{eqn:ke2kmu2}
\eeq
This result is in good agreement with the SM expectation and has a relative error
($\sim 1.3\%$) three times smaller compared to the previous world average.
Further improvements in the knowledge of $(R_K^{e/\mu})_{exp.}$ would be more than
welcome. Moreover, also the KLOE collaboration will reach an error down to the 1\%
level on $R_K^{e/\mu}$, once the remaining statistics will be added and
the reconstruction efficiency improved~\cite{flavianet}.

Last but not least, an error on $(R_K^{e/\mu})_{exp.}$ of about $0.3\%$ is the
ambitious goal of the 2007 dedicated run of the CERN-P326 collaboration
(the successor of NA48)~\cite{flavianet}.
If these expectations will be fulfilled, in a short term the error on the 
world average of $R_K^{e/\mu}$ will decrease by an additional factor of four.

In the following, we consider low-energy minimal SUSY extensions of the SM (MSSM)
with R parity as the source of NP to be tested by $R_K^{e/\mu}$~\cite{kl2lfv}.
As discussed in \cite{kl2lfv}, it is indeed possible for regions of the
MSSM to obtain $\Delta r^{e-\mu}_{\!NP}$ of $\mathcal{O}(10^{-2})$ and,
such large contributions to $K_{\ell2}$, do not arise from SUSY lepton
flavor conserving (LFC) effects, but, rather, from lepton flavor violating
(LFV) ones.

\begin{figure}
    \begin{center}
    \raisebox{48mm}{      \begin{tabular}{lc}
        \hline \hline
        & $(R_K^{e/\mu})_{exp.}$ $[10^{-5}]$  \\ \hline
        PDG 2006~\cite{PDG}                       & $2.45 \pm 0.11$ \\
        NA48/2~'03~prel.   & $2.416 \pm 0.043 \pm 0.024$ \\
        NA48/2~'04~prel.   & $2.455 \pm 0.045 \pm 0.041$ \\
        KLOE prel.            & $2.55 \pm 0.05 \pm 0.05$ \\ \hline
        SM prediction                             & $2.472 \pm 0.001$ \\
        \hline \hline
      \end{tabular}}
     \vskip -2.5 cm 
     \caption{Current experimental data on $R_K^{e/\mu}$ from \cite{flavianet}.
     \label{fig:susylimit} }
   \end{center}
\end{figure}
The main reason is that, whenever new physics acts in $K_{\ell 2}$ to create a
departure from the strict SM $\mu-e$ universality, these new contributions will
typically be proportional to the lepton masses.
Hence, what occurs in the SUSY case is that LFC contributions are suppressed
with respect to the LFV ones by higher powers of the first two generations
lepton masses (it turns out that the first contributions to $\Delta r^{e-\mu}_{\!NP}$
from LFC terms arise at the cubic order in $m_{\ell}$, with $\ell=e,\mu$).
Instead, for the LFV contributions to $R_K^{e/\mu}$ one can select those which involve 
flavor changes from the first two lepton generations to the third one with the possibility
of picking up terms proportional to the tau-Yukawa coupling which can be large in
the large $\rm{\tan\beta}$ regime (the parameter $\rm{\tan\beta}$ denotes the ratio
of Higgs vacuum expectation values responsible for the up- and down- quark masses,
respectively). Moreover, the relevant one-loop induced LFV Yukawa interactions are
known \cite{bkl} to acquire an additional $\rm{\tan\beta}$ factor with
respect to the tree level LFC Yukawa terms.
Thus, the loop suppression factor can be (partially) compensated in the
large $\rm{\tan\beta}$ regime.

In this paper, we analise the domain of $\Delta r_K^{e/\mu}$ between $10^{-2}$
and $10^{-3}$. We show that:

i) if  $\Delta r_K^{e/\mu}$ is found to be  $\Delta r_K^{e/\mu}\geq 5\times 10^{-3}$, 
then the signal unambiguously indicates the presence of LFV sources.

ii) if  $\Delta r_K^{e/\mu}\leq 5\times 10^{-3}$, then both the LFC and LFV
sources can account for the effect. 

iii) a value of $\Delta r_K^{e/\mu}$ between $5\times 10^{-3}$ and $10^{-3}$ severely
constrains the parameter space in the $M_H-\tan\beta$ plane.

iv) if a signal exists at a such a level, the LHC results become the crucial tool
to discriminate between the LFC and LFV  sources of LFU breaking.

v) there exists a strong correlation between large LFU violation and LFV in lepton 
  decays (mainly $\tau$ decays); another interesting relation concerns the regions 
  of SUSY parameter space where the deviation from the SM expectation for the muon
  anomalous magnetic moment finds a SUSY explanation and that allowing for a sizeable 
  LFU violation.

The paper is organized as follow: in Section~\ref{LFU_general}, we outline general
considerations about LFU in $P_{l2}$. In section Section~\ref{LFU_LFV}, we specialize
to the LFV case while in Section~\ref{LFU_LFC}, we discuss the additional possibility
of LFC contributions.
In Section~\ref{constraints}, we list the constraints we have imposed on the SUSY
parameter space before starting the analysis of the LFU breaking effects.
In Section~\ref{correlations}, we discuss the correlation between LFU violation
and LFV in lepton decays and their possible connection with a SUSY explanation
for the anomalous magnetic moment of the muon.
In Section~\ref{N_analysis}, we present the quantitative analysis of our results
incorporating the constraints of the above sections.
In Section~\ref{2HDM}, we extend the analysis of LFU breaking effects to a generic
two Higgs Doublet Model with tree level flavor changing interactions between the
Higgs bosons and the fermions.
Finally, in Section~\ref{conclusion} we summarize the main results of the present
analysis.

\section{Lepton Flavor Universality in $P_{\ell 2}$}
\label{LFU_general}

Within the SM, possible departures from the LFU are predicted to be
\begin{equation}
  |\Delta r^{\ell_1/\ell_2}_{\rm SM}| =
  {\cal O}[(\alpha/4\pi)\times(m^{2}_{\ell_{1(2)}}/M^{2}_{W})]\,,
\end{equation}
and thus completely negligible. This explains why the study of LFU
breaking represents a very useful tool to look for NP effects.

On general grounds, violations of LFU in charged current interactions can be
classified into two classes: i) corrections to the strength of the effective
 $(V-A)\times(V-A)$ four-fermion interaction, ii) four-fermion interactions
with new Lorentz structures.

As an example of the first class, we mention the $W\ell\nu_{\ell}$ vertex correction
through a loop of new particles: the induced effect is of order 
$(\alpha/4\pi)\times (M^{2}_{W}/M^{2}_{NP})$,
hence unobservably small.
Second class is definitely more promising: the typical example is the
scalar current induced by tree level Higgs exchange, with mass-dependent
coupling ($H\ell\nu \sim m_{\ell}\tan\beta$). 

In the following, we will analyze LFU breaking effects arising from this latter 
class occurring in $P_{\ell 2}$.

Due to the V-A structure of the weak interactions, the SM contributions
to $P_{\ell2}$ are helicity suppressed; hence, these processes are very
sensitive to non-SM effects (such as multi-Higgs effects) which might
induce an effective pseudoscalar hadronic weak current.

In particular, charged Higgs bosons ($H^\pm$) appearing in any model with
two Higgs doublets (including the SUSY case) can contribute at tree level
to the above processes.

The relevant four-Fermi interaction for the decay of charged mesons induced
by $W^\pm$ and $H^\pm$ has the following form:
%
%
%
\beq
\label{EH}
\frac{4G_F}{\sqrt{2}}V_{ud}
\left[(\,\overline{u}\gamma_{\mu}P_Ld\,)
(\,\overline{l}\gamma^{\mu}P_L\nu_{l}\,) +
\Delta^{ij}\,t_{\beta}^{2}\left(\frac{m_{d} m_{l_i}}{m^{2}_{H^\pm}}\right)
(\,\overline{u}P_Rd\,)(\,\overline{l}_iP_L\nu_j\,)\right]\,,
\eeq
where $P_{R,L}=(1\pm \gamma_5)/2$ and we kept only the $t_{\beta}$
(with $t_{\beta} = \tan\beta$) enhanced part of the $H^\pm ud$ coupling,
namely the $m_d t_{\beta}$ term.

The quantity $\Delta^{ij}=\Delta^{ij}(\delta^{ij},\tan\beta,m_{l_j},\tilde{m})$
may depend, in general, on the mixing angle $\delta^{ij}$ regulating the flavor
transition $ij$, on the $\tan\beta$ parameter, on the masses of all charged lepton
generations and, finally, on all the possible susy masses $\tilde{m}$ determining
the effective vertex.

The decays $P\rightarrow \ell\nu$ ($P=K,\pi$) proceed via the axial-vector part
of the $W^\pm$ coupling and via the pseudoscalar part of the $H^\pm$ coupling.
Then, once we implement the PCAC's
\beq
<0|\overline{u}\gamma_{\mu}\gamma_{5}d|M^{-}>=if_M p^{\mu}_M\,\,\,\,\,\,,\,\,\,\,\,
<0|\overline{u}\gamma_{5}d|M^{-}>=-if_M \frac{m^{2}_M}{m_{d}+m_u}\,,
\eeq
it is found that
\begin{equation}
R_{P\ell_i\nu}=
\bigg[
1-\Delta^{ii}\bigg(\frac{m_{d_P}}{m_{d_P}\!+\!m_{u_P}}\bigg)
\frac{m^{2}_{P}}{M^{2}_{H^+}}t_{\beta}^{2}
\bigg]^2
+
\Sigma_{j\neq i}|\Delta^{ij}|^2
\bigg(\frac{m_{d_P}}{m_{d_P}\!+\!m_{u_P}}\bigg)^2
\frac{m^{4}_{P}}{M^{4}_{H^+}}t_{\beta}^{4}
\label{tree}
\end{equation}
The tree level charged Higgs exchange leads to a contribution with $i=j$
and $\Delta^{ii}=1$.
However, the introduction of a charged scalar current (induced by a $H^+$) does not
introduce any deviation from the SM expectation of the LFU breaking in $R_P^{e/\mu}$.

Indeed, we observe that the SM amplitude is proportional to $m_\ell$ because of the
helicity suppression while the charged Higgs one carries the $m_\ell$ dependence
through the Yukawa coupling. 

As a result, the first SUSY contributions violating the $\mu-e$ universality in
$P\to\ell\nu$ decays arise at the one-loop level with various diagrams involving
 exchanges of (charged and neutral) Higgs scalars, charginos, neutralinos and sleptons.
For our purpose, it is relevant to divide all such contributions into two classes:

i) LFC contributions, where the charged meson M decays without FCNC in the leptonic
sector, i.e. $P\to\ell\nu_{\ell}$;

ii) LFV contributions $P\to\ell_i\nu_k$, with $i$
and $k$ referring to different generations (in particular, the interesting case will
be for $i= e,\mu$, and $k=\tau$).

In the following sections we address separately the case of LFC and LFV contributions.

\section{The lepton flavor violating case}
\label{LFU_LFV}

Within SUSY theories, there exist two different classes of LFV interactions:
\begin{enumerate}

\item[i)] Gauge-mediated LFV interactions\,,
\item[ii)] Higgs-mediated LFV interactions\,.
\end{enumerate}
As regards the class $i)$, LFV effects are induced by the exchange
of gauginos and sleptons; these contributions decouple with the
heaviest mass $m_{SUSY}$ circulating in the slepton/gaugino loops.

Concerning the case $ii)$, we remind that models containing at least two Higgs 
doublets generally allow flavor violating couplings of the Higgs bosons with the fermions~\cite{Gunion:1989we}. However in the MSSM such LFV couplings are absent 
at tree level since we have one higgs doublet coupling uniquely to the up-sector,
while the other higgs doublet couples only to the down-sector.
However, once non holomorphic terms are generated by loop effects (so called HRS
corrections~\cite{Hall:1993gn}) and given a source of LFV among the sleptons,
Higgs-mediated $H\overline{\ell}_i\ell_j$ LFV couplings are unavoidable~\cite{bkl}.
These effects decouple with the heavy Higgs mass scale $m_H$ but they do not decouple
with the mass scale of the sleptons/gauginos circulating in the loop, given that the
effective LFV Yukawa couplings arise from dimension four operators. As it is well known,
higgs mediated effects to rare decays start being competitive with the gaugino mediated 
ones when $m_{SUSY}$ is roughly one order of magnitude heavier then $m_H$ and for
$\tan\beta\sim\mathcal O(50)$ \cite{Paradisi:2005tk}.
On general ground, there is no reason to assume that $m_H\simeq m_{SUSY}$, unless specific
models of SUSY breaking are assumed.

We stress that the quantity which is determined experimentally and accounts for
the deviation from the $\mu-e$ universality is
\beq
(R_P^{e/\mu})_{exp.}=
\frac{\sum_i\Gamma(P\rightarrow e\nu_i)}
{\sum_i\Gamma(P\rightarrow \mu\nu_{i})}
\,\,\,\,\,\,\,\,\,\,\,\,\,\,\,\,\,\,\,\,\,\,\,\,\,\,\,\,i= e,\mu,\tau.
\eeq
with the sum extended over all (anti)neutrino flavors. In fact, experimentally,
it is possible to measure only the charged lepton flavor in the decay products.

The dominant SUSY contributions to $R_P^{e/\mu}=(R_P^{e/\mu})_{exp.}/(R_P^{e/\mu})_{SM}$
arise from the charged Higgs exchange.

One could naively think that the SUSY effects in the LFV channels $P\to\ell_i\nu_k$
are further suppressed with respect to the LFC ones. On the contrary, charged Higgs
mediated LFV contributions, in particular in the kaon decays into an electron
or a muon and a tau neutrino, can be strongly enhanced.

In particular, the expressions for the effective couplings $\Delta^{ij}$ in Eq.~\ref{EH} read
\bea
\label{LFCcoupl}
\Delta^{\ell\ell} &=&
\frac{1}{(1+\epsilon t_{\beta})(1+\epsilon_{\ell}t_{\beta})} +
\frac{m_{\tau}}{m_{\ell}}\frac{\Delta^{\ell\ell}_{RL}t_{\beta}}
{(1+\epsilon t_{\beta})(1+\epsilon_{\tau}t_{\beta})^2}\\
\label{LFVcoupl}
\Delta^{\ell\tau} &=&
\frac{m_{\tau}}{m_{\ell}}
\frac{\Delta^{3l}_{R}t_{\beta}}{(1+\epsilon t_{\beta})(1+\epsilon_{\tau}t_{\beta})^2}
\qquad\qquad\qquad\qquad\qquad\qquad\qquad l=e,\mu\,.
\eea
The first term in Eq.~\ref{LFCcoupl} refers to a tree level charged Higgs exchange
while the second one stems from a double source of LFV that, as a final result,
preserves the flavor.
On the contrary, the contributions of Eq.~\ref{LFVcoupl} refer to LFV channels.
Notice that the (loop induced) contributions arising from LFV sources is enhanced
by the factor $m_{\tau}/m_{\ell}$ (compared to the contributions from a tree level
charged Higgs exchange) when the electron or muon in $(R_P^{e/\mu})_{exp.}$
are accompanied by a tau neutrino.

In the above expressions, we have also included the threshold corrections
(proportional to $\epsilon, \epsilon_{\ell}$, with $\epsilon\sim\alpha_{s}/4\pi$
and $\epsilon_{\ell}\sim \alpha_{2}/4\pi$)) for the quark and lepton yukawas
appearing when we integrate out heavy degrees of freedom from the low energy
effective theory \cite{Hall:1993gn}.

In particular, a relevant observation for the following analysis is that the one-loop
induced $\epsilon_{\ell}=\epsilon_{\ell}(m^{2}_{\tilde{\ell}},M^{2}_{\tilde{\chi}})$
resummation factors carry a lepton flavor dependence through the slepton masses.
Thus, as we will see in the next section, if the slepton generations have different
masses, the $\epsilon_{\ell}$ factors will generate a breaking of the LFU in low-energy
observables.

The $\Delta^{3\ell}_{R, RL}$ terms are induced at one loop level by the exchange of
Bino or Bino-Higgsino and sleptons.
Since the Yukawa operator is of dimension four, the quantities $\Delta^{3\ell}_{R}$
depend only on ratios of SUSY masses, hence avoiding SUSY decoupling. In the so called
mass insertion (MI) approximation, the expressions of $\Delta^{3\ell}_{R, RL}$ 
are given by:
\beq
\Delta^{3\ell}_{R}\!\simeq\! \frac{\alpha_{Y}}{8\pi}\mu M_1 m^{2}_{R} 
\delta^{3\ell}_{RR}
\left[I^{'}\!(M^{2}_{1},\mu^2,m^{2}_{R})\!-\!(\mu\!\leftrightarrow\! m_{L})
\right]
\label{Delta_R}
\eeq
\beq
\Delta^{\ell\ell}_{RL}\!\simeq\!
-\frac{\alpha_{Y}}{16\pi}\mu M_1 m^{2}_{L} m^{2}_{R}
\,\delta^{\ell 3}_{RR}\delta^{3\ell}_{LL}\,I^{''}\!(M^{2}_{1},m^{2}_{L},m^{2}_{R})\,,
\label{Delta_RL}
\eeq
where $\mu$ is the the Higgs mixing parameter, $M_1$ is the Bino ($\tilde{B}$) mass
and $m^{2}_{L(R)}$ stands for the left-left (right-right) slepton mass matrix entry.
The LFV MIs, i.e.
$\delta^{3\ell}_{XX}\!=\!({\tilde m}^2_{\ell})^{3\ell}_{XX}/m^{2}_{X}$ $(X=L,R)$,
are the off-diagonal flavor changing entries of the slepton mass matrix.
The loop function $I^{'}(x,y,z)$ is such that $I^{'}(x,y,z)= dI(x,y,z)/d z$,
where $I(x,y,z)$ refers to the standard three point one-loop integral which
has mass dimension -2; morever, $I^{''}(x,y,z)= d^2I(x,y,z)/dydz$.
As it is clearly shown by Eq.~\ref{Delta_R}, $\Delta^{3\ell}_{R}$ vanishes for
$\mu=m_L$. On the other hand, both $\Delta^{3\ell}_{R}$ and $\Delta^{3\ell}_{RL}$
assume the their maximum values when $\mu \gg M_1,m_L,m_R$; this is easily 
understood reminding that Higgs mediated effects originate from non holomorphic 
corrections that are driven by the $\mu H_1H_2$ term in the superpotential.

In particular, in the limit where $\mu \gg \tilde{m}\!=\!M_1\!=\!m_L\!=\!m_R$, 
it turns out that $\Delta^{3\ell}_{R}\simeq\alpha_{Y}/16\pi\times \mu/\tilde{m}\times\delta^{3\ell}_{RR}$ and $\Delta^{\ell\ell}_{RL}\simeq\alpha_{Y}/32\pi\times \mu/\tilde{m}\times
\delta^{\ell 3}_{RR}\delta^{3\ell}_{LL}$
\footnote{Im($\delta^{13}_{RR}\delta^{31}_{LL}$) is strongly constrained by the
electron electric dipole moment \cite{Masina:2002mv}.
However, sizable contributions to $R^{LFV}_{K}$ can still be induced by
Re($\delta^{13}_{RR}\delta^{31}_{LL}$).}.
Making use of the effective couplings of Eq.~(\ref{LFVcoupl}), it turns out that
the dominant contribution to $\Delta r^{e-\mu}_{NP}$ reads
\begin{eqnarray}
\label{LFUlfv}
R^{e/\mu}_{K}\simeq
\left|1-\frac{m^{2}_{K}}{M^{2}_{H}}
\frac{m_{\tau}}{m_{e}}
\frac{\Delta^{11}_{RL}t_{\beta}^{3}}{(1+\epsilon t_{\beta})
(1+\epsilon_{\tau}t_{\beta})^2}\right|^{2}+
\bigg(\frac{m^{4}_{K}}{M^{4}_{H}}\bigg)
\bigg(\frac{m^{2}_{\tau}}{m^{2}_{e}}\bigg)
\frac{|\Delta^{31}_{R}|^2t_{\beta}^{6}}
{(1+\epsilon t_{\beta})^2(1+\epsilon_{\tau}t_{\beta})^4}.
\end{eqnarray}
In the above expression, we have included the interference between SM and
SUSY LFC terms (arising from a double LFV source). 
In Eq.~(\ref{LFUlfv}) terms proportional to $\Delta^{32}_{R}$ are neglected
given that they are suppressed by a factor $m^{2}_{e}/m^{2}_{\mu}$ with
respect to the term proportional to $\Delta^{31}_{R}$.

Taking $\Delta^{31}_{R}\!\simeq\!5\cdot 10^{-4}$ (by means of a numerical
analysis, it turns out that $\Delta^{3\ell}_{R}\leq 10^{-3}$ 
\cite{Brignole:2003iv}), $\tan\beta\!=\!40$ and $M_{H}\!=\!500 GeV$ we end
up with $\Delta r^{e-\mu}_{\!K\,SUSY}\simeq 10^{-2}$.
We see that in the large (but not extreme) $\rm\tan\beta$ regime and with
a relatively heavy $H^{\pm}$, it is possible to reach contributions to
$\Delta r^{e-\mu}_{\!K\,SUSY}$ at the percent level thanks to the possible
LFV enhancements arising in SUSY models.

Turning to pion physics, one could wonder whether the analogous quantity
$\Delta r^{e-\mu}_{\!\pi\,SUSY}$ is able to constrain SUSY LFV.
However,  the correlation between $\Delta r^{e-\mu}_{\!\pi\,SUSY}$ and
$\Delta r^{e-\mu}_{\!K\,SUSY}$:
\beq
\label{lfvpi}
\Delta r^{e-\mu}_{\pi\,SUSY}\simeq\left(\frac{m_d}{m_u+m_d}\right)^{2}
\left(\frac{m^{4}_{\pi}}{m^{4}_{k}}\right)
\Delta r^{e-\mu}_{\!K\,SUSY}
\eeq
clearly shows that the constraints on $\Delta r^{e-\mu}_{\!K\,susy}$
force $\Delta r^{e-\mu}_{\pi\,susy}$ to be much below its current
experimental upper bound.

\section{The lepton flavor conserving case}
\label{LFU_LFC}

We now reconsider Eq.~\ref{EH} in the $i=j$ case, i.e.
the lepton flavor conserving channels. In absence of LFV interactions, 
$\Delta_{ii}$ reads
$\Delta_{ii} = 1/[(1+\epsilon t_{\beta})(1+\epsilon_{\ell}t_{\beta})]$. This leads to
\begin{equation}
\frac{\Gamma(P\to\ell\nu)}{\Gamma(P\to\ell\nu)_{SM}}
= \bigg[1-\bigg(\frac{m_{d_P}}{m_{d_P}\!+\!m_{u_P}}\bigg)
\frac{m^{2}_{P}}{M^{2}_{H^+}}
\frac{t_{\beta}^{2}}{(1+\epsilon t_{\beta})(1+\epsilon_{\ell}t_{\beta})}
\bigg]^2
\label{tree}
\end{equation}
As discussed in the previous section, tree level $H^+$ contributions do not introduce
any breaking of the LU in $R_P^{e/\mu}$. However, this is strictly true only if 
$\epsilon_{e}=\epsilon_{\mu}$, as clearly shown by Eq.~\ref{tree}. In particular,
for non universal slepton masses, it turns out that $\epsilon_{e}\neq\epsilon_{\mu}$
(remind that $\epsilon_{\ell}=\epsilon_{\ell}(m^{2}_{\tilde{\ell}},M^{2}_{\tilde{\chi}})$),
and LFU breaking effects are generated. By means of Eq.\ref{tree}, we find that
$\Delta r_P^{e/\mu}$ is well approximated by the following expression
\begin{equation}
\Delta r_P^{e/\mu}\simeq
-2\,\bigg(\frac{m_{d_P}}{m_{d_P}\!+\!m_{u_P}}\bigg)
\frac{m^{2}_{P}}{M^{2}_{H^+}}
\frac{t_{\beta}^{3}}{(1+\epsilon t_{\beta})}(\epsilon_{e}-\epsilon_{\mu})\,,
\label{PlnuLU}
\end{equation}
where we have assumed that $m^{2}_{P}/M^{2}_{H^+}\ll 1$ and $\epsilon_{\ell}t_{\beta}\ll 1$.
We observe that the NP sensitivity to the above effects of $K\to\ell\nu$ is higher than that 
of $\pi\to\ell\nu$ by a factor of
$\Delta r_K^{e/\mu}/\Delta r_{\pi}^{e/\mu} \sim m^{2}_{K}/m^{2}_{\pi}$.
The current experimental resolutions on these modes imply that $K\to\ell\nu$ is the
best probe of the above scenario.

Moreover, LFC contributions to $R_{\pi}$ and $R_{K}$ can be also induced  at the loop level
by box, wave function renormalization and vertex contributions from SUSY particle exchange
~\cite{kl2lfv}. The complete calculation of the $\mu$ decay in the 
MSSM~\cite{Chankowski:1993eu,Krawczyk:1987zj}
can be easily applied to the meson decays. The dominant contributions to $\Delta r^{e-\mu}_{SUSY}$
arise from  the charginos/neutralinos sleptons ($\tilde l_{e,\mu}$) exchange and it has the form
~\cite{kl2lfv}
\beq
\Delta r^{e-\mu}_{SUSY}\sim \frac{\alpha_{2}}{4\pi}
\left(\frac{\tilde{m}^{2}_{\mu}-\tilde{m}^{2}_{e}}{\tilde{m}^{2}_{\mu}+
\tilde{m}^{2}_{e}}\right)\frac{m^{2}_{W}}{M^{2}_{SUSY}},
\eeq
thus, $\Delta r^{e-\mu}_{SUSY}$ can be of order $\Delta r^{e-\mu}_{SUSY}\leq 10^{-3}$, provided
there exists a large mass splitting among sleptons ($\tilde{m}^{2}_{\mu}\ll\tilde{m}^{2}_{e}$
or $\tilde{m}^{2}_{\mu}\gg\tilde{m}^{2}_{e}$) and a SUSY mass scale $M_{SUSY}$ not much above the
EW scale, i.e. $M_{SUSY} \sim m_{W}$. So, it turns out that all these LFC contributions yield
values of $\Delta r^{e-\mu}_{\!K\,SUSY}$ which are smaller than the current and expected future
experimental sensitivities in kaon physics.

On the other hand, given that the NP sensitivity to the above effects of $\Delta r_K^{e/\mu}$ and
$\Delta r_{\pi}^{e/\mu}$ is the same and since the experimental resolution is better in the pion
system, for this flavor conserving SUSY contribution it is the decay $\pi\to\ell\nu$ to represent
the best place where to look for LFU violation. In particular,
\beq
R^{exp.}_{\pi}=
(1.230\pm 0.004)\cdot 10^{-4}\,\,\,\,\,\,\,\rm{PDG}
\eeq
and by making a comparison with the SM prediction
\beq
R^{SM}_{\pi}=(1.2354\pm 0.0002)\cdot 10^{-4}
\eeq
one obtains (at the $2\sigma$ level) 
\beq
-0.0107\leq\Delta r^{e-\mu}_{\!NP}\leq 0.0022.
\eeq
Comparing this interval for $\Delta r^{e-\mu}_{\!NP}$ with the above value of 
$\Delta r^{e-\mu}_{SUSY}$,  it turns out that only under rather particular
conditions (very large mass splitting of sleptons of different generation,
relatively light SUSY scale) can one obtain visible LFC SUSY contributions
to the LFU violation in pion decays \cite{musolf}.

\section{Constraints}
\label{constraints}

In this section, we list the constraints we have imposed on
the SUSY parameter space before starting the analysis of the
LU breaking effects.

\subsection{Direct SUSY search}
\label{sect:susysearch}

The framework in which we work is a low-energy R-parity conserving susy model
with generic LFV soft breaking terms.
We perform a scan up to a mass scale of $5\rm{TeV}$ of the following low energy 
parameters: the gaugino masses $M_i$ ($i=1,3$), the $\mu$ term, the left-left
and right-right sfermion mass terms for the first two and the third generations 
$M_{\tilde{f}}$, the trilinear coupling in the stop sector $A_t$; moreover
$\tan\beta< 60$. At the low scale, we impose the following constraints on each point:
\begin{itemize}
\item Lower bound on the light and pseudo--scalar Higgs masses~\cite{Higgs_ALEPH};
\item The LEP constraints on the lightest chargino and sfermion masses~\cite{Eidelman:2004wy};
\item The LEP and Tevatron constrains on squarks, gluino and charged Higgs
masses~\cite{Eidelman:2004wy}
\item Absence of charge and/or colour breaking minima~\cite{CaDimo}.
\item The lightest susy particle (LSP) is neutral.
\item Electroweak Precision Observables (EWPO) constraints~\cite{EWPO}.
\end{itemize}
For future convenience, it is useful to recall which are the necessary
conditions under which the lightest Higgs mass bounds are satisfied. 
First of all, we remind that the LEPII bound $(m_{h})_{\rm SM}\gsim 114$ GeV,
applies also to SUSY theories, irrespective of the $\tan\beta$ values,
provided we assume the decoupling regime, roughly implying that
$M_{H,A}\geq 200 GeV$. Indeed, this will represent the case under study 
in the present work. Within the MSSM,
the lightest Higgs mass is bounded from above. In particular, we can write
$m_h^2=m_h^{2({\rm tree})}+m_h^{2({\rm loop})}$ where, for large $\tan\beta$,
$m_h^{2({\rm tree})}\sim m_Z^2-4m_Z^2m_A^2/(m_A^2-m_Z^2)\cot^2\beta$.
The most significant loop contribution is given by
\bea
m_h^{2({\rm loop})}&=&\frac{3m_t^4}{4\pi^2
  v^2}\left[\ln\left(\frac{m_{\tilde t_1}m_{\tilde t_2}}{m_t^2}\right)
  +\frac{|X_t|^2}{m_{\tilde t_1}^2-m_{\tilde t_2}^2}
  \ln\left(\frac{m_{\tilde t_1}^2}{m_{\tilde t_2}^2}\right)\right.\no\\
&&\left.+\frac12\left(\frac{|X_t|^2}{m_{\tilde t_1}^2-m_{\tilde
        t_2}^2}\right)^2
  \left(2-\frac{m_{\tilde t_1}^2+m_{\tilde t_2}^2}{m_{\tilde
        t_1}^2-m_{\tilde t_2}^2}
    \ln\left(\frac{m_{\tilde t_1}^2}{m_{\tilde
          t_2}^2}\right)\right)\right],
\eea
where $X_t=A_t-\mu^*\cot\beta$.
Thus, the tree level contribution, that is maximum for moderate to large
$\tan\beta$, has to be supplemented by sizable loop corrections.
In particular, if the stop mixing is small, $|X_t/m_{\tilde t_{1,2}}|^2\ll1$,
the correction depends only on the logarithm of the stop masses, so these
must be rather heavy. If, however, the stop mixing is large, much lighter
stops can still yield large loop corrections. However, as we will see, this
last possibility is disfavored by the $b\to s\gamma$ constraints, specially
in the large $\tan\beta$ regime.

\subsection{B-physics observables}
\label{sect:low}

\subsubsection{\boldmath $\BR(B\rightarrow X_s \gamma)$}

As it is well known, $\BR(B\rightarrow X_s \gamma)$ is a particularly sensitive
observable to possible non-standard contributions and it provides a non-trivial
constraint on the SUSY mass spectrum given its precise experimental determination
and the very accurate SM calculation at the NNLO~\cite{bsgth}.
According to the recent NNLO analysis of Ref.~\cite{bsgth}, the SM prediction is
${\cal B}(B\to X_s \gamma; E_\gamma > 1.6~{\rm GeV})^{\rm SM}
= (3.15 \pm 0.23) \times 10^{-4}$.
Combining this result with the experimental average~\cite{hfag,belle,babar}
${\cal B}(B\to X_s \gamma; E_\gamma > 1.6~{\rm GeV}))^{\rm exp}
 = (3.55 \pm 0.24) \times 10^{-4}$ we obtain
\bea
R_{ Bs\gamma} =
 \frac{ {\cal B}^{\rm exp}(B\to X_s \gamma)}
{{\cal B}^{\rm SM}(B\to X_s \gamma)}
= 1.13\pm 0.12\,.
\label{eq:Bsgamma}
\eea
In our numerical analysis we impose the above constraint at the $2\sigma$ C.L.

Within Minimal Flavor Violating (MFV) frameworks \cite{MFV}, the dominant SUSY
contributions to $\BR(B\rightarrow X_s \gamma)$ arise from the one-loop charged-Higgs
and chargino-squark amplitudes. Charged-Higgs effects unambiguously increase the rate
compared to the SM expectation, while the chargino-squark ones can have both signs
depending on the sign of $sign(\mu A_{\tilde{t}})$.
In this work we choose $\mu>0$ (that is also preferred by the $(g-2)_{\mu}$ constraints)
and $sign(A_{\tilde{t}})<0$, which implies destructive interference between chargino
and charged Higgs contributions.
A simple expression accounting for NP contributions in $B\to X_s\gamma$ is provided
by \cite{bsgth,Misiak}
\beq
R_{ Bs\gamma} \simeq 1 - 2.54\,C^{NP}_{7}(M_W) - 0.60\,C^{NP}_{8}(M_W)
\label{bsg}
\eeq
where $C^{NP}_{7,8}\!=\!C^{\tilde{\chi}^\pm}_{7,8} + C^{H^\pm}_{7,8}$ are the relevant
Wilson coefficients for the New Physics contributions evaluated at the $M_W$ scale.
In particular, starting from the full expressions for $C^{NP}_{7,8}$of Ref.~\cite{bsgamma},
one can derive the following approximate expressions 
\bea
  C^{H^\pm}_{7} &\simeq&
  \left(\frac{1-\epsilon\,t_{\beta}}{1+\epsilon\,t_{\beta}}\right)
  \,\frac{m^2_t}{M^{2}_{H^\pm}}\,
  F^{7}_{H^\pm}\left(\frac{m^2_t}{M^{2}_{H^\pm}}\right)\,,
  \nonumber \\
  C^{\tilde{\chi}^\pm}_{7} &\simeq&
  -\frac{A_{\tilde{t}}}{\mu}\,\frac{m^2_t}{\mu^2}\,
  \frac{t_{\beta}}{1+\epsilon\,t_{\beta}}\,
  F^{7}_{\tilde{\chi}^\pm}\left(\frac{m_{\tilde{q}}^{2}}{\mu^2}\right)\,,
\eea
where $\epsilon\sim 10^{-2}$ for a degenerate SUSY spectrum and
$F^{7}_{\tilde{\chi}^\pm}(1)\simeq 0.07$,
$F^{7}_{\tilde{\chi}^\pm}(x\gg 1)\simeq (13/12-1/2\log(x))/x^2$ and
$F^{7}_{\tilde{\chi}^\pm}(x\ll 1)\simeq 7/12+2/3\log(x)$
while $F^{7}_{H^\pm}(x_{tH})\simeq 1/4 + 1/3\log (x_{tH})$ for
$x_{tH}=m^2_t/M^{2}_{H^\pm}\ll 1$ and $F^{7}_{H^\pm}(1)\simeq -0.2$.
We observe that, when $m_{\tilde{q}}/\mu \ll 1$, the lower bound on
$m_{\tilde{q}}$ is set by the experimental limits on the lightest
stop mass $m^{2}_{\tilde{t}_1}\simeq m^{2}_{\tilde{q}}-m_t|A_{\tilde{t}}|$
and on the sbottom mass $m^{2}_{\tilde{b}_1}\simeq m^{2}_{\tilde{q}}-m_b|\mu|t_{\beta}$.
Similar expressions for the subleading contributions proportional to $C^{NP}_{8}$
are not shown, although included in our numerical analysis.

Taking $C^{H^{\pm}}_{7}$ and $C^{\tilde{\chi}^{\pm}}_{7}$ separately,
the following observations follow: i) the lower bound on $m_{H^\pm}\geq 295 \rm{GeV}$,
holding at the $2\sigma$ level within a 2HDM framework ( where it is assumed
$\epsilon = 0$ and where $C^{\tilde{\chi}^\pm}_{7}=0$), can be significantly
relaxed within SUSY scenarios thanks to a reduction of $C^{H^{\pm}}_{7}$
driven by the threshold corrections $\epsilon$;
in particular if $\tan\beta\sim 50$ it turns out that $m_{H^\pm}\geq 200\rm{GeV}$
ii) for a natural scenario where all the SUSY masses have comparable size,
in particular for $A_{\tilde{t}}/(\mu,m_{\tilde{q}})\sim 1$, the regime
$\tan\beta\sim 50$ necessarily implies that $\mu$ and/or $m_{\tilde{q}}$ lie in
the $\geq 1\rm{TeV}$ scale iii) the simultaneous requirement of large values for 
$\tan\beta$ and relatively light $m_{\tilde{q}},\mu$ ( below the $ 1\rm{TeV}$ scale) 
necessarily implies either large cancellations between 
$C^{H^\pm}_{7}$ and $C^{\tilde{\chi}^\pm}_{7}$ and/or 
$A_{\tilde{t}}/(\mu,m_{\tilde{q}})$ significantly less than 1. However, as we have 
seen before, the scenario with relatively light 
$m_{\tilde{q}}$ and $A_{\tilde{t}}/m_{\tilde{q}}\ll 1$ is constrained by the lower
bound on the lightest Higgs mass $m_h$.

\subsubsection{\boldmath $B\to\tau\nu$}

The recent Belle\cite{Btaunu_Belle} and BaBar\cite{Btaunu_Babar} results
for $B\to\ell\nu$ leads to the average ${\cal B}(\Btaun)^{\rm exp}=(1.42\pm 0.43)\times 10^{-4}~$. This should be compared with the SM expectation
${\cal B}(\Btaun)^{\rm SM}=G_{F}^{2} m_{B}m_{\tau}^{2}f_{B}^{2}|V_{ub}|^{2}
(1-m_{\tau}^{2}/m_{B}^{2})^{2}\tau_B/8\pi$, whose numerical value
suffers from sizable parametrical uncertainties induced by $f_B$ and $V_{ub}$.
Taking $\tau_B = (1.643\pm 0.010)\rm{ps}$, $V_{ub} = (4.00\pm 0.26)\times 10^{-3}$
and $f_B = 0.216\pm 0.022\rm{GeV}$ \cite{fB}, the best estimate is
${\cal B}(\Btaun)^{\rm SM} = (1.33\pm 0.23)\times 10^{-4}~$, which
implies
\bea
R_{B\tau\nu}^{\rm exp} =
\frac{\BR^{\rm exp}(\Btaun)}{\BR^{\rm SM}(\Btaun)}
~=~ 1.07 \pm 0.37~.
\label{eq:Rtn_exp}
\eea

From the theoretical side, the $B\to\ell\nu$ process is one of the cleanest
probes of the large $\tan\beta$ scenario due to its enhanced sensitivity to
tree-level charged-Higgs exchange \cite{hou,tgbhints}. In particular, a scalar
charged current induced by NP theories with extended Higgs sectors, provides
the following effects:
\begin{equation}
R_{B\ell\nu}= \bigg[1-\frac{m^{2}_{B}}{M^{2}_{H^+}}
\frac{t_{\beta}^{2}}{(1+\epsilon t_{\beta})(1+\epsilon_{\ell}t_{\beta})}
\bigg]^2
\label{Plnu}
\end{equation}
where we have included corrections both for the quark and lepton yukawas
arising within SUSY theories.

The new physics effect on 
$R_{K\mu \nu}=\Gamma^{\rm SUSY}(K\to\mu\nu)/\Gamma^{\rm SM}(K\to\mu\nu)$ is 
obtained from Eq.~\ref{Plnu} with the replacement $m_B^2\to m_K^2$
\cite{tgbhints}. Although the charged Higgs contributions are now suppressed by
a factor $m_K^2/m_B^2\simeq 1/100$, this is well compensated by the excellent 
experimental resolution \cite{flavianet} and the good theoretical 
control.
However, given that these new physics effects are, in the most favorable
cases, at the \% level, we would need a theoretical prediction for the SM contribution
at the same level to use this decay as an effective constraint. We would then need 
an independent determination both of $f_K$ (possibly from lattice QCD) and $V_{us}$
with such a level of accuracy.

The best strategy to fully exploit the New Physics sensitivity of $K_{l2}$ systems
is to consider the ratio $R^{'}=R_{K\mu\nu}/R_{\pi\mu\nu}$ \cite{tgbhints,flavianet}.
In fact, on the one side $R^{'}$ and $R_{K\mu\nu}$ have the same New Physics content
(being $R_{\pi\mu\nu}$ not sizably affected by charged current interactions) and,
on the other side, $R^{'}$ depends on $(f_K/f_{\pi})^2$ instead of $f_K^2$ with 
$f_K/f_{\pi}$ being much better under control compared to $f_K$ by means of lattice QCD.
However, at present, unquenched lattice calculations of $f_K/f_{\pi}$ are still 
not well established. Therefore, although it may play a relevant role in the future, 
we do not include the constraints from $K\to l\nu$ in the present analysis.

The above argument for $K\to l\nu$ does not apply to $B\to\ell\nu$. In fact, even if
the hadronic uncertainties related to $f_B$ and $V_{ub}$ are much larger that those
for $f_K$ and $V_{us}$, they cannot hide in any way the huge NP effects in 
$B\to\ell\nu$ that can arise in our scenario.

\subsubsection{\boldmath $B_{s}\rightarrow \mu^{+} \mu^{-}$}
\label{sec:bsmumu}

The SM prediction for ${\mathcal B}(B_{s}\rightarrow \mu^{+} \mu^{-})$ is
${\mathcal B}(B_s\rightarrow\mu^{+}\mu^{-})_{{\rm SM}}=(3.37\pm 0.31)\times 10^{-9}$.
This value should be compared to the present 95\% C.L. upper bound from
CDF, ${\rm BR}(B_s\rightarrow \mu^{+} \mu^{-})_{\rm exp} < 5.8 \times 10^{-8}$
that still leaves a large room for NP contributions. In particular, the MSSM
with large $\tan\beta$ allows, in a natural way, large differences between SM and
SUSY expectations for ${\mathcal B}(B_{s}\rightarrow\mu^{+}\mu^{-})$ \cite{bkq}.
The SUSY contributions can be summarized by the approximate formula
%
%
%
\beq 
{\rm BR}(B_s\to\mu^+\mu^-)\simeq
\frac{5\times10^{-8}}{\left[1+0.5 \times\frac{\tan \beta}{50}\right]^4}
\Bigg[\frac{\tan\beta}{50}\Bigg]^6
\left(\frac{500 \rm{GeV}}{M_A}\right)^4
\left(\frac{\epsilon_{Y}}{3\times 10^{-3}}\right)^{2}
\label{bsmumuSUSY}
\eeq
where $\epsilon_{Y}\simeq-1/16\pi^2\times A_t/\mu \times H_2(y_{u_R},y_{u_L})$
with $y_{q_{R,L}}=M^2_{\tilde{q}_{L,R}}/|\mu|^2$, $H_2(1,1)=-1/2$,
$H_2(x\gg1,y=x)\simeq -1/x$ and $H_2(x\ll 1,y=x)\simeq 1+\log x$;
thus, $\epsilon^{\tilde{\chi}^-}_{Y} \sim 3\times 10^{-3}$ holds
in the limit of all the SUSY masses and $A_t$ equal.
As we can see, the present CDF upper bound on ${\mathcal B}(B_{s}\to\mu^{+}\mu^{-})$
already provides constraints in some regions of the SUSY parameter space.
Moreover, we remind that, although also $\Delta M_s$ is a New Physics sensitive
observable in the scenario we are considering, it doesn't provide any further
constraints in addition to those inferred by the B-physics observables that we
have already discussed.

\subsection{\boldmath $(g-2)_{\mu}$}

The possibility that the anomalous magnetic moment of the muon [$a_\mu = (g-2)_{\mu}/2$],
which has been measured very precisely in the last few years \cite{g_2_exp}, provides a
first hint of physics beyond the SM has been widely discussed in the recent literature.
Despite substantial progress both on the experimental and on the theoretical sides, the
situation is not completely clear yet (see Ref.~\cite{g_2_th} for an updated discussion).

Most recent analyses converge towards a $3\sigma$ discrepancy in the $10^{-9}$ range~\cite{g_2_th}:
\beq
 \Delta a_{\mu} =  a_{\mu}^{\rm exp} - a_{\mu}^{\rm SM}
\approx (3 \pm 1) \times 10^{-9}~.
\label{eq:amu_exp}
\eeq
Recently, Passera et al.~\cite{passera_mh} have considered the possibility that the present
discrepancy between experiment and the Standard Model (SM) prediction for $(g-2)_{\mu}$ may
arise from errors in the determination of the hadronic leading-order contribution to the latter.
If this is the case, the authors of Ref.~\cite{passera_mh} find a decrease on the electroweak
upper bound on the SM Higgs boson mass. By means of a detailed analysis they conclude that
this solution of the muon $(g-2)_{\mu}$ discrepancy is unlikely in view of current experimental
error estimates.

The main SUSY contribution to $a^{\rm MSSM}_\mu$ is usually
provided by the loop exchange of charginos and sneutrinos.
The basic features of the supersymmetric contribution to $a_\mu$
are correctly reproduced by the following approximate expression:
\beq
\frac{a^{\rm MSSM}_\mu}{ 1 \times 10^{-9}}  
\approx 1.5\left(\frac{\tan\beta }{10} \right) 
\left( \frac{300~\rm GeV}{m_{\tilde \nu}} \right)^2
\left(\frac{\mu M_2}{m^{2}_{\tilde \nu}} \right)~,
\label{eq:g_2}
\eeq
which provides a good approximation to the full one-loop result
\cite{g_2_mw}.

The most relevant feature of Eqs.~(\ref{eq:g_2}) is that the sign of
$a^{\rm MSSM}_\mu$ is fixed by the sign of the $\mu$ term so that the
$\mu>0$ region is strongly favored.

\section{LFU vs LFV and the $(g-2)_{\mu}$ anomaly}
\label{correlations}

As we have previously seen, sizable LFU breaking effects can be generated in SUSY 
through LFV interactions which involve the third generation. Hence, a legitimate
worry is whether the bounds on LFV tau decays, like $\tau\rightarrow eX$ 
(with $X=\gamma,\eta,\mu\mu$), are respected in the region of SUSY parameter space
leading to a strong enhancement of the LFU violation \cite{Paradisi:2005tk}.
The present and projected bounds (to be achieved at a super B factory) on some of
these processes are summarized in Table \ref{tab:exp}.
\begingroup
 \begin{table}
 \begin{center}
 \begin{tabular*}{0.8\textwidth}{@{\extracolsep{\fill}}||c|c|c|c||}
 \hline\hline
&Process & Present Bounds & Expected Future Bounds  \\[0.2pt] 
 \hline
 (1) &  BR($\tau \to e,\gamma$) & $9.4~ \times~ 10^{-8}$ &
 $\mathcal{O}(10^{-8}) $ \\
 (2) &  BR($\tau \to e,e,e$) & $2.0~ \times~ 10^{-7}$ &
 $\mathcal{O}(10^{-8}) $ \\
 (3) &  BR($\tau \to e,\mu,\mu$) & $2.0~ \times~ 10^{-7}$ &
 $\mathcal{O}(10^{-8}) $ \\
 (4) &  BR($\tau \to e,\eta$) & $4.5~ \times~ 10^{-8}$ &
 $\mathcal{O}(10^{-8}) $ \\
\hline\hline
 \end{tabular*}
 \end{center}
 \caption{Present and Upcoming experimental limits on various
leptonic processes at 90\% C.L.}
 \label{tab:exp}
 \end{table}
 \endgroup
The most sensitive probe of Higgs mediated effects is generally provided by
$\tau\rightarrow \ell_j\eta$~\cite{sher}; the corresponding branching ratio
is given by~\cite{brignolerossi,herrero}
\beq
\frac{Br(\tau\rightarrow l_j\eta)}{Br(\tau\rightarrow l_j\bar{\nu_j}\nu_{\tau})} 
\simeq 18\pi^2\!\left(\frac{f^{8}_{\eta} m^{2}_{\eta}}{m_{\tau}}\right)^{\!2}\!
\!\left(1\!-\!\frac{m^{2}_{\eta}}{m^{2}_{\tau}}\right)^{\!2}
\left(\frac{|\Delta^{3j}|^2t_{\beta}^{6}}{m^{4}_{A}}\right)
\eeq
where $m^{2}_{\eta}/m^{2}_{\tau}\simeq 9.5\times10^{-2}$ and the relevant decay constant
is $f^{8}_{\eta}\sim 110 {\rm MeV}$.
Moreover, $|\Delta^{3j}|^2\!=\!|\Delta^{3j}_{L}|^2+|\Delta^{3j}_{R}|^2$, where $\Delta^{3j}_{L}$
has a similar expression to $\Delta^{3j}_{R}$~\cite{brignolerossi} and it is such that $\Delta^{3j}_{L}\sim\delta^{3j}_{LL}$. We note that, in order to generate a non-vanishing
$\Delta r^{e-\mu}_{\!K\,Susy}$, RR-type flavor structures are unavoidable; on the contrary,
$Br(\tau\rightarrow e\eta)$ can be generated by both LL and/or RR-type mixing angles, being
$Br(\tau\rightarrow e\eta)\sim|\Delta^{31}_{L}|^2+|\Delta^{31}_{R}|^2$.
Given that $\Delta r^{e-\mu}_{\!K\,Susy}$ and $Br(\tau\rightarrow e \eta)$
have the same SUSY dependence, the upper bound on $Br(\tau\rightarrow eX)$
is automatically found once we saturate the allowed range (at the \% level)
for New Physics contributions in $\Delta r^{e-\mu}_{\!K\,Susy}$. We find that
\beq
Br(\tau\rightarrow e\eta)
\simeq
10^{-2}\left(\frac{|\Delta^{31}|^2t_{\beta}^{6}}{m^{4}_{A}}\right)
\simeq 10^{-8}\times\Delta r^{e-\mu}_{\!K\,Susy}\,,
\label{rktmueta}
\eeq
where the last equality holds when $\Delta^{31}_{L}=0$. So, employing the
constraints for $\Delta r^{e-\mu}_{\!K\,Susy}$ at the $\%$ level, we obtain
$Br(\tau\rightarrow e\eta)\leq 10^{-10}$.
We conclude that, the present and expected experimental upper bounds
on $Br(\tau\rightarrow e\eta)$ (see Table~\ref{tab:exp}) still
allow large effects in $\Delta r^{e-\mu}_{\!K\,Susy}$.

On the other hand, $\tau\rightarrow\ell_j\gamma$ is the most sensitive probe
of LFU violation induced by SUSY gauge mediated effects.

In contrast to the $Br(\tau\rightarrow e\eta)$ case, it is not possible
to link $Br(\tau\rightarrow e\gamma)$ and $\Delta r^{e-\mu}_{\!K\,Susy}$
in a way that is independent of the specific choice for the susy breaking sector.
In particular, as discussed before, the New Physics contributions
to $\Delta r^{e-\mu}_{\!K\,Susy}$ decouple with the heavy Higgs mass $m_H$,
while $Br(\tau\rightarrow e\gamma)$ decouples with the heaviest SUSY particle
mass $\tilde{m}$ circulating in the gaugino/slepton loop.

In the following, to get a feeling of where we stand, we will evaluate $Br(\tau\rightarrow e\gamma)$
in the region of the parameter space where large LFU breaking effects in $\Delta r^{e-\mu}_{\!K\,Susy}$
can be generated.
In particular, a necessary ingredient in order to get large $\Delta r^{e-\mu}_{\!K\,Susy}$
values is to maximize the size of the effective LFV coupling $\Delta^{31}_{R}$ (remember
that $\Delta r^{e-\mu}_{\!K\,Susy}\sim t^{6}_{\beta}/M^{4}_{H}\times \Delta^{31}_{R}$) and
this happens when $\mu\gg\tilde{m}$. In this latter case, starting from the full expressions
of Ref.~\cite{hisano_95}, we find the following expression
%
%
 \beq
 \frac{BR(\tau\rightarrow\ell_{j}\gamma)}
 {BR(\tau\rightarrow\ell_{j}\nu_{\tau}\bar{\nu_j})} \simeq
 \frac{\pi\alpha_{el}}{3G_{F}^{2}}\,
\left(\frac{\alpha_{Y}}{4}\right)^{2}
\left( \left|\delta_{LL}^{3j}\right|^2 + \left|\delta_{RR}^{3j}\right|^2 \right)
\,\frac{\mu^2}{\tilde{m}^2}\,\frac{t^{2}_{\beta}}{\tilde{m}^4}\,.
\label{tauegamma1_bis}
 \eeq
From Eq.~\ref{tauegamma1_bis} we can get
 \beq
 BR(\tau\rightarrow \ell_j\gamma)\approx
 5\times 10^{-8}
 \left(\,
\left|\frac{\delta_{RR}^{3j}}{0.5}\right|^2 +
 \left|\frac{\delta_{LL}^{3j}}{0.5}\right|^2
\,\right)\,
\left(\frac{t_{\beta}}{50}\right)^2\,
\left(\frac{1\,\rm{TeV}}{\tilde{m}}\right)^4\,
\frac{\mu^2}{\tilde{m}^2}\,,
\label{tauegamma1_bis2}
 \eeq
showing that large (order one) mixing angles for $\delta_{LL,RR}^{3j}$ are phenomenologically 
allowed, provided there exists a rather heavy spectrum for the soft sector. For instance, for $\mu/\tilde{m}=4$, it turns out that $\tilde{m}\geq 2 \rm{TeV}$. Obviously, such a lower bound
on $\tilde{m}$ can be relaxed for smaller values of $\delta_{LL,RR}^{3j}$ and/or $\tan\beta$.
Moreover, one can easily find the following approximate expression
\beq
\Delta r^{e-\mu}_{\!K\,Susy}
\leq
10^{-1}\times
\frac{BR(\tau\rightarrow e\gamma)}{10^{-7}}
\left(\frac{\tilde{m}/M_{H}}{4}\right)^{4}
\left(\frac{t_{\beta}}{50}\right)^{4}
\,,
\label{rkvsteg}
 \eeq
showing that, in the large $\tan\beta$ regime and for heavy Higgs masses lighter
than those of the soft breaking terms, experimentally visible LFU breaking effects
in $K\to\ell \nu$ can be naturally obtained.
In particular, large LFU breaking effects even above the $10\%$ level (already
excluded experimentally), can be always compatible with the experimental constraints
on $BR(\tau\rightarrow e\gamma)$ for slepton/gaugino masses at the $\rm{TeV}$ scale.
However, we stress again that it is not possible to correlate $BR(\tau\to e\gamma)$
to $\Delta r^{e-\mu}_{\!K\,Susy}$, unless specific SUSY breaking mechanisms (relating
$\tilde{m}$ and $M_{H}$) are assumed.

On the contrary, the processes $\ell_i\rightarrow \ell_j\gamma$ are intimately
linked to the muon anomalous magnetic moment $(g-2)_{\mu}$, as they both arise
from dipole transitions \cite{hisano_gm2}.

Thus, in the following, we address the interesting question of whether it is
possible, within SUSY theories, to account for the current $(g-2)_{\mu}$ anomaly,
while generating, at the same time, LFU breaking effects in $\Delta r^{e-\mu}_{\!K\,Susy}$
at the $\%$ level. As we will show, the answer is positive and this will lead to
set a lower bound on $BR(\tau\rightarrow e\gamma)$.
To see this point explicitly, let us derive the correlation between
$BR(\ell_i\rightarrow \ell_j\gamma)$ and $(g-2)_{\mu}$ for the relevant
case where $\mu/\tilde{m}\gg 1$; in this case, $\Delta a_{\mu}$ is well
approximated by the expression
\bea
  \Delta a_{\mu}&\simeq&
\frac{\alpha_Y}{24\pi}\,\frac{\mu}{\tilde{m}}\,
\frac{m_{\mu}^{2}}{\tilde{m}^{2}}\,t_{\beta}\nonumber\\
&\simeq& 
3\times 10^{-9}\, \left(\frac{\mu/\tilde{m}}{5}\right)\,
\left(\frac{400\rm{GeV}}{\tilde{m}}\right)^2\,
\left(\frac{t_{\beta}}{50}\right)
\label{gm2}
\eea
and thus we find that
\bea
BR(\tau\rightarrow \ell_j\gamma)
&\simeq&
\frac{12\pi^3}{m^{4}_{\mu}}\,
\left(\frac{\alpha_{el}}{G^{2}_{f}}\right)
\left(\Delta a_{\mu}\right)^2
\left(
      \left|\delta_{RR}^{3j}\right|^2 +
      \left|\delta_{LL}^{3j}\right|^2
\right)
BR(\tau\rightarrow \ell_j\nu_{\tau}\bar{\nu}_{j})
\nonumber\\
&\simeq&
 3 \times 10^{-9}\,
\left(\frac{\Delta a_{\mu}}{1 \times 10^{-9}}\right)^2\,
\left(\,
\left|\frac{\delta_{RR}^{3j}}{0.01}\right|^2 +
\left|\frac{\delta_{LL}^{3j}}{0.01}\right|^2
\,\right)\,.
\label{tauegammagm2}
\eea
From Eqs.~\ref{rkvsteg}, \ref{gm2}, \ref{tauegammagm2} we conclude that, LFU
breaking effects at the $\%$ level, typically implying $\delta_{RR}^{31}\geq 0.01$,
are compatible with the current experimental bounds on $BR(\tau\rightarrow e\gamma)$;
moreover, if we additionally require that SUSY effects explain the current
discrepancy for the muon anomalous magnetic moment, i.e.
$\Delta a_{\mu}\geq1\times10^{-9}$ at the $2\sigma$ level \cite{g_2_th}, LFU
breaking effects at the $\%$ level unavoidably imply large effects in 
$BR(\tau\rightarrow e\gamma)\geq 3\times 10^{-9}$, well within the expected
reach of a superB factory.


As we have seen, sizable LFU breaking effects originating from LFV interactions
require a flavor mixing in the $13$ sector. Thus, from a phenomenological point
of view, $\Delta r^{e-\mu}_{\!K\,Susy}$ is naturally related to $\tau-e$
transitions as, for instance, $\tau\rightarrow e\eta$, $\tau\rightarrow e\gamma$
etc. However, a legitimate question that can be addressed is what one would expect
for $\tau-\mu$ and $\mu-e$ transitions when sizable sources of LFV in the $\tau-e$
sector are assumed.

In particular, from a model building point of view, it seems hard to generate
large effects for $\tau-e$ transitions while keeping the effects for $\tau-\mu$
transitions small. Moreover, once $\tau-\mu$ transitions are induced, an effective $(\mu-e)_{eff.}$ transition of the type $(\mu-e)_{eff.}=(\mu-\tau)\times(\tau-e)$
is also induced and processes like $\mu\to e\gamma$ are unavoidable.

In the following, we will address the above issue more quantitatively.
In particular, the analogue expression of Eq.~\ref{tauegamma1_bis} for
the $\mu\rightarrow e\gamma$ case reads
 \bea
 \frac{BR(\mu\rightarrow e\gamma)}
 {BR(\mu\rightarrow e\nu_{\mu}\bar{\nu_e})} &=&
 \frac{\pi\alpha_{el}}{3G_{F}^{2}}\,
\frac{t^{2}_{\beta}}{\tilde{m}^4}
\left(\frac{\alpha_{Y}}{10}\right)^{2}
\left|
\delta_{RR}^{23}\delta_{RR}^{31}+
\frac{m_{\tau}}{m_{\mu}}\delta_{LL}^{23}\delta_{RR}^{31}
\right|^2
\left(\frac{\mu}{\tilde{m}}\right)^{2}
\,,
\label{muegamma}
 \eea
where, besides the combination of LFV sources relevant for the present discussion,
i.e. $RR$-type LFV sources, we have also kept the $\delta_{LL}^{23}\delta_{RR}^{31}$
contribution.
In fact, this last contribution is enhanced, at the amplitude level, by the ratio
$m_{\tau}/m_{\mu}$ compared to the $\delta_{RR}^{23}\delta_{RR}^{31}$ contribution
and thus, potentially large even when $\delta_{LL}^{23}\!<\!\delta_{RR}^{32}$.
Finally, it turns out that
 \beq
 BR(\mu\rightarrow e\gamma)\simeq 10^{-11}
\left|
\frac{\delta_{RR}^{23}\delta_{RR}^{31}}{10^{-2}}+
\frac{m_{\tau}}{m_{\mu}}
\frac{\delta_{LL}^{23}\delta_{RR}^{31}}{10^{-2}}
\right|^2\,
\left(\frac{t_{\beta}}{50}\right)^2\,
\left(\frac{1\,\rm{TeV}}{\tilde{m}}\right)^4
\left(\frac{\mu}{\tilde{m}}\right)^{2}
\,.
\label{muegamma2}
 \eeq
Eq.~\ref{muegamma2} shows that it is not possible ( in the large $\tan\beta$ regime, at least)
to have simultaneously order one MIs $\delta_{RR}^{23}$ and $\delta_{RR}^{31}$, unless we push
$\tilde{m}$ in the multi-TeV regime. In particular, if $\delta_{RR}^{23}=\delta_{RR}^{31}=0.1$, $\delta_{LL}^{23}=0$ and $\mu/\tilde{m}=4$, it turns out that $\tilde{m}\geq 2 \rm{TeV}$.
Clearly, such a scenario is not compatible with an explanation of the $(g-2)_{\mu}$ anomaly.

Moreover, as we have discussed in the previous sections, there is also the
possibility to obtain negative values for $\Delta r^{e-\mu}_{\!K\,Susy}$
(see Eq.~\ref{LFUlfv}) when both RR and LL-type of flavor violating sources
for the $1-3$ transition are present. However, this possibility can be constrained,
in some cases, by the experimental upper bounds on $BR(\mu\rightarrow e\gamma)$.
In fact, combining Eq.~\ref{LFUlfv} with Eq.~\ref{muegamma}, it turns out that
\beq
|\Delta r^{e-\mu}_{\!K\,Susy}|
\leq
3 \times 10^{-3}\times
\sqrt{\frac{BR(\mu\rightarrow e\gamma)}{10^{-11}}}
\left(\frac{\tilde{m}/M_{H}}{10}\right)^{2}
\left(\frac{t_{\beta}}{50}\right)^{2}
\left|\frac{\delta^{31}_{LL}}{\delta^{32}_{LL}}\right|
\,,
\label{muegamma_rk}
 \eeq
thus, unless we assume $\delta^{32}_{LL}/\delta^{31}_{LL}\!<<\!1$ (that is typically unnatural
from a model building point of view), we are lead with large effects in $BR(\mu\to e\gamma)$
even for an heavy soft sector at the $\rm{TeV}$ scale.
However, we stress that from a pure phenomenological perspective, $BR(\mu\rightarrow e\gamma)$
doesn't impose a direct bound on $\Delta r^{e-\mu}_{\!K\,Susy}$ given that different parameters
enter the two quantities.

Let us finally point out that, when $\delta_{RR}>>\delta_{LL}$, the following upper bound on
$\Delta r^{e-\mu}_{\!K\,Susy}$ holds
\beq
\Delta r^{e-\mu}_{\!K\,Susy}\leq
\frac{Br(\tau\rightarrow \mu\eta)}{10^{-8}}
\times
\frac{BR(\tau\rightarrow e\gamma)}{BR(\tau\rightarrow \mu\gamma)}\,.
\label{rk_tmueta_teg_tmug}
\eeq
Clearly, only the discovery of LFV signals in some of the above observables,
by means of improved experimental sensitivities, would shed light on the
scenarios outlined before.

We conclude this section pointing out that the same New Physics effect
observable in the Kaon system through $\Delta r^{e-\mu}_{\!K\,Susy}$
is also observable, in principle, in B physics systems by means of purely
leptonic decays of charged $\rm{B}$ meson. In particular, it is found that
\beq
\frac{BR(B\to e\nu)}{BR(B\to e\nu)_{SM}}=
\left[1 + \frac{m^{4}_{B}}{m^{4}_{K}}\,
\Delta r^{e-\mu}_{\!K}\right]
\simeq
\left[1+10^{2}\times\left(\frac{\Delta r^{e-\mu}_{\!K}}{10^{-2}}\,\right)\right]\,.
\label{rktmueta}
\eeq
This means that, a LFU breaking effect at the $\%$ level in the $K\ell2$ systems,
implies an enhancement of two orders of magnitude in $BR(B\to e\nu)$ compared
to its SM expectation. However, given that $BR(B\to e\nu)_{SM}\approx 10^{-11}$,
an experimental sensitivity at the level of $BR(B\to e\nu)_{exp.}\leq 10^{-9}$
would be necessary.


\section{Numerical Analysis}
\label{N_analysis}

In the following, we will analyze the allowed size for the LFU breaking effects
in $R^{e/\mu}_{P}$ both in the lepton flavor conserving and violating cases.

\begin{figure}
\includegraphics[scale=0.38]{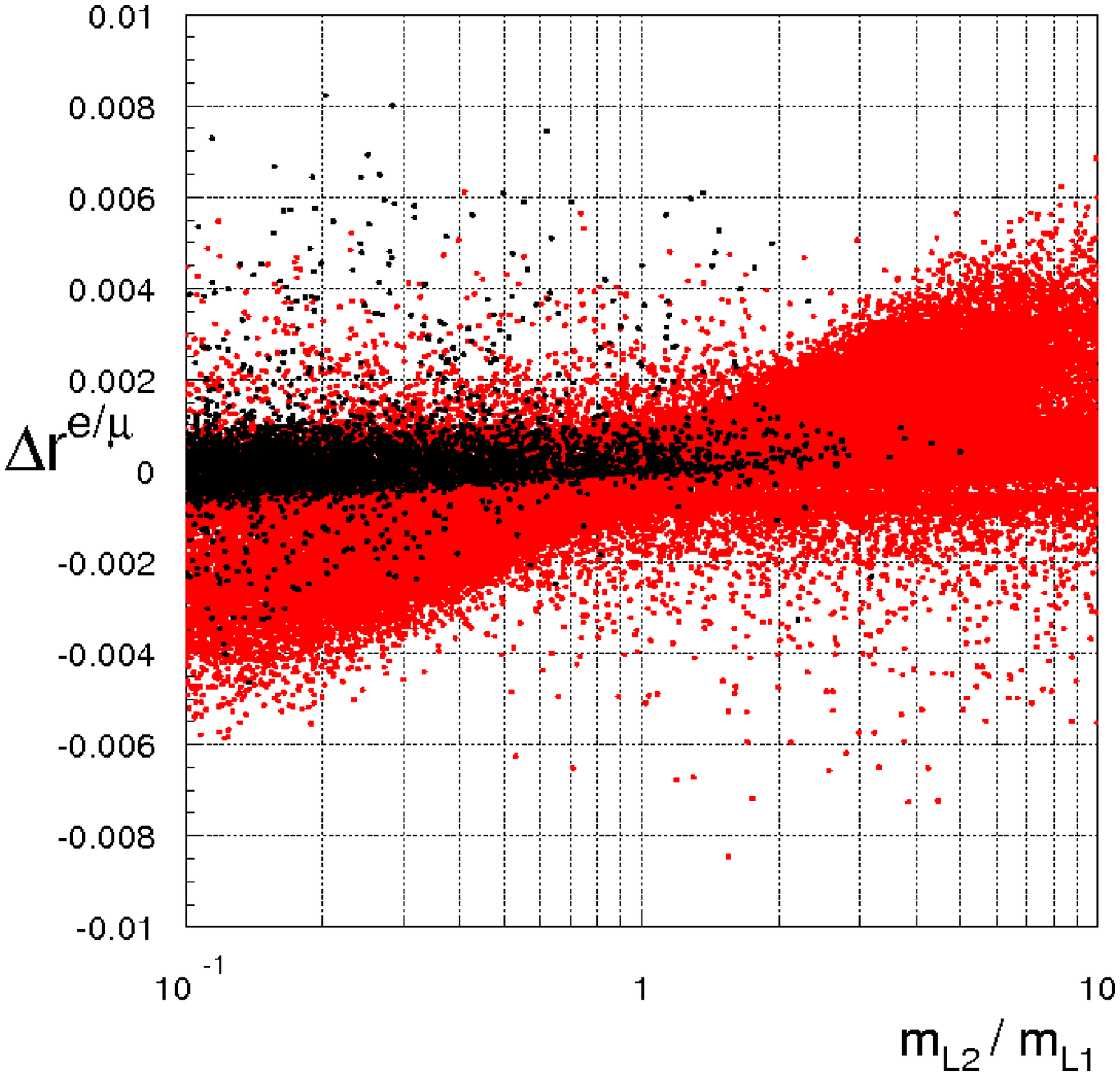}
\includegraphics[scale=0.38]{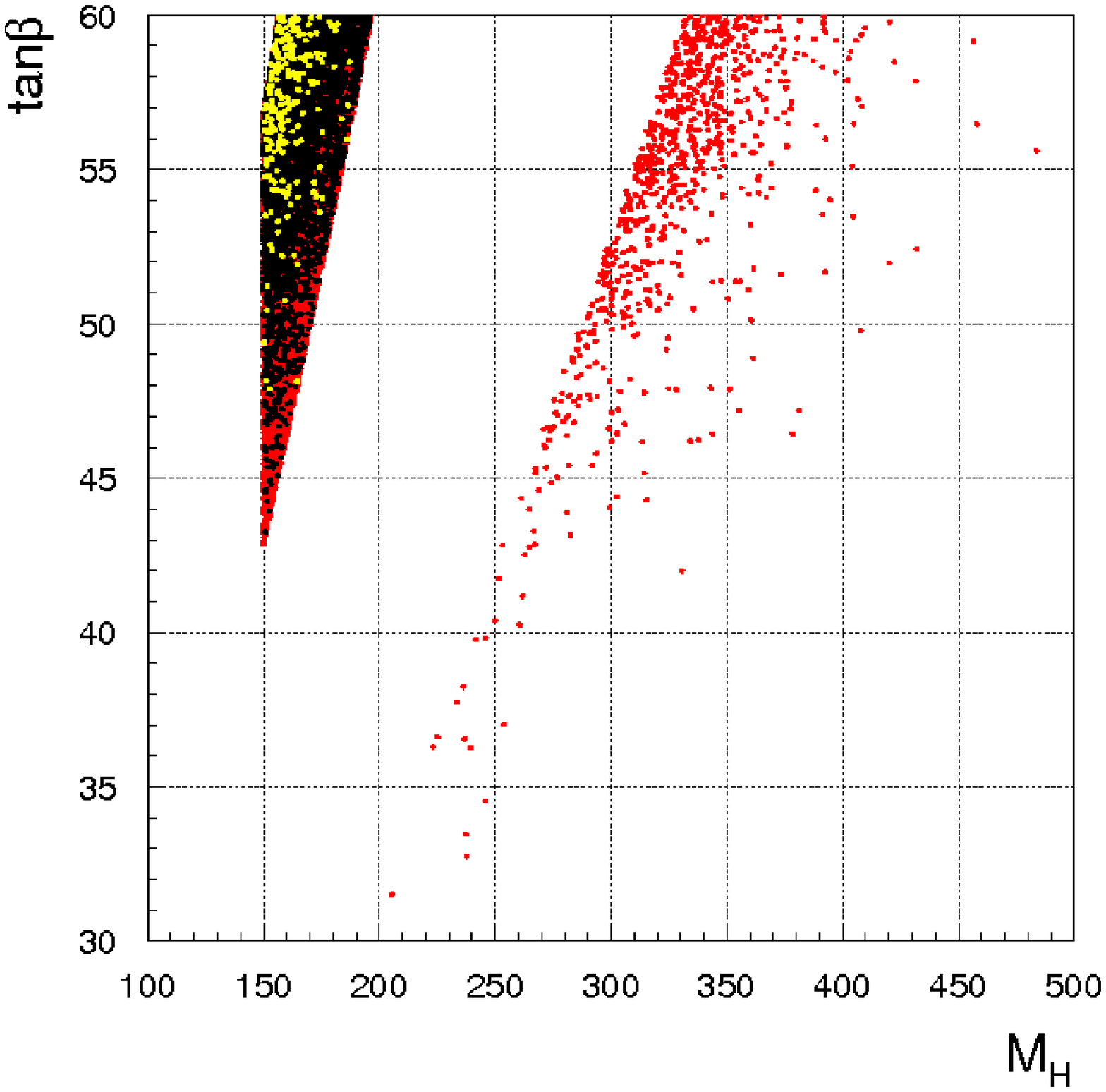}
\caption{\label{LUlfc}
SUSY Lepton Flavor Conserving contributions to $\Delta r_K^{e/\mu}$.
Left: $\Delta r_K^{e/\mu}$ as a function of the (left-handed) mass ratio
between the second and the first slepton generations. Black dots refer to
the points satisfying $1\times 10^{-9}<\Delta a_\mu <5\times 10^{-9}$.
Right: regions of the parameter space in the $\tan\beta-M_H$ plane where
$0.001\!<\!|\Delta r_K^{e/\mu}|\!<\! 0.003$ (red dots),
$0.003\!<\!|\Delta r_K^{e/\mu}|\!<\! 0.005$ (black dots)
and $|\Delta r_K^{e/\mu}|\!>\! 0.005$ (yellow dots).
The plot has been obtained by means of a scan over the following parameter
space: $(m_{L_{1,2}}, m_{R_{1,2}}, m_{\tilde{Q}}, m_{\tilde{g}}, m_{\tilde{W}},
m_{\tilde{B}}, M_{H})<2.5\rm{TeV}$, $\mu\!<\!5$\rm{TeV} and $\tan\beta<60$.
All the dots present in these and subsequent figures satisfy all the constraints
discussed in Sec.~\ref{constraints}.}
\end{figure}

In the former case, LFU breaking effects arise from mass splittings between
sleptons of the first and second families $(m_{L1(L2)}$, $m_{R1(R2)}$), as 
discussed in Sec.~\ref{LFU_LFC}.
In Fig.~\ref{LUlfc}, we perform a numerical analysis of the allowed values for
$\Delta r_K^{e/\mu}$ (see Eq.\ref{PlnuLU}) through a scan over the following 
SUSY parameter space: $(m_{L1(L2)}$, $m_{R1(R2)}$, $m_{\tilde{Q}}$, $m_{\tilde{g}}$,
$m_{\tilde{W}}$, $m_{\tilde{B}}$, $M_{H})<2.5\rm{TeV}$, $\mu\!<\!5$\rm{TeV} and
$\tan\beta<60$.
In particular, we allow different entries for the left-left (LL) and the right-right
(RR) blocks in the slepton mass matrix for the first two generations, i.e. for
$m_{L1(L2)}$ and $m_{R1(R2)}$ respectively.
Moreover, we also impose all the constraints discussed in Sec.~\ref{constraints}.
In Fig~\ref{LUlfc}, on the left, we show $\Delta r_K^{e/\mu}$ as a function of
the (left-handed) mass splitting between the second and first slepton generations.
Black dots refer to the points satisfying the $(g-2)_\mu$ discrepancy at the $95\%$
C.L., i.e. $1\times 10^{-9}<\Delta a_\mu <5\times 10^{-9}$.

As we can see, the maximum LFU breaking effects are reached for maximum mass 
splitting between sleptons. However, when $m_{L1}\!=\!m_{L2}$,
we would expect LFU breaking effects going to zero, in contrast to what is 
shown by Fig~\ref{LUlfc}. This happens because mass splittings for right-handed 
sleptons $m_{R1}\!\neq\!m_{R2}$ (not explicitly visible in Fig~\ref{LUlfc}), 
can still generate LFU breaking effects even in the case where $m_{L1}\!=\!m_{L2}$.
We see that values for $|\Delta r_K^{e/\mu}|$ as large as $5\times 10^{-3}$ 
are possible for slepton masses splitted by a factor  $10$.
However, potentially visible values for $|\Delta r_K^{e/\mu}|$ of order of
$\sim 2\times 10^{-3}$ are obtained even for smaller mass splittings, i.e.
for $m_{L1,L2}/m_{L2,L1}\sim 2$.

Interestingly enough, the sign of these LFU breaking effects depends on the ratio
between the slepton masses. In particular, if the left-handed smuons are heavier
then the selectrons $\Delta r_K^{e/\mu}>0$, while $\Delta r_K^{e/\mu}<0$ if the
smuons are lighter then the selectrons. The opposite situation happens for mass
splittings of right-handed smuons.

In Fig~\ref{LUlfc}, on the right, we show the regions of the parameter space
in the $\tan\beta-M_H$ plane where $0.001\!<\!|\Delta r_K^{e/\mu}|\!<\! 0.003$
(red dots), $0.003\!<\!|\Delta r_K^{e/\mu}|\!<\! 0.005$ (black dots) and
$|\Delta r_K^{e/\mu}|\!>\! 0.005$ (yellow dots).
We observe that the narrow region where $m_H\leq 200$\rm{GeV} corresponds to the
points where the $B\to\tau\nu$ constraints are not effective. This does not occur
not because the new physics contributions to $B\to\tau\nu$ are small; quite the
contrary, the reason is that they are quite large ($\sim 2\times$ SM ones) and
destructively interfere with the SM contribution (see Eq.~\ref{Plnu}).
On the other hand, the region of the $\tan\beta-M_H$ plane between the two
allowed areas is excluded by the $B\to\tau\nu$ constraint.

\begin{figure}
\centering
\includegraphics[scale=0.38]{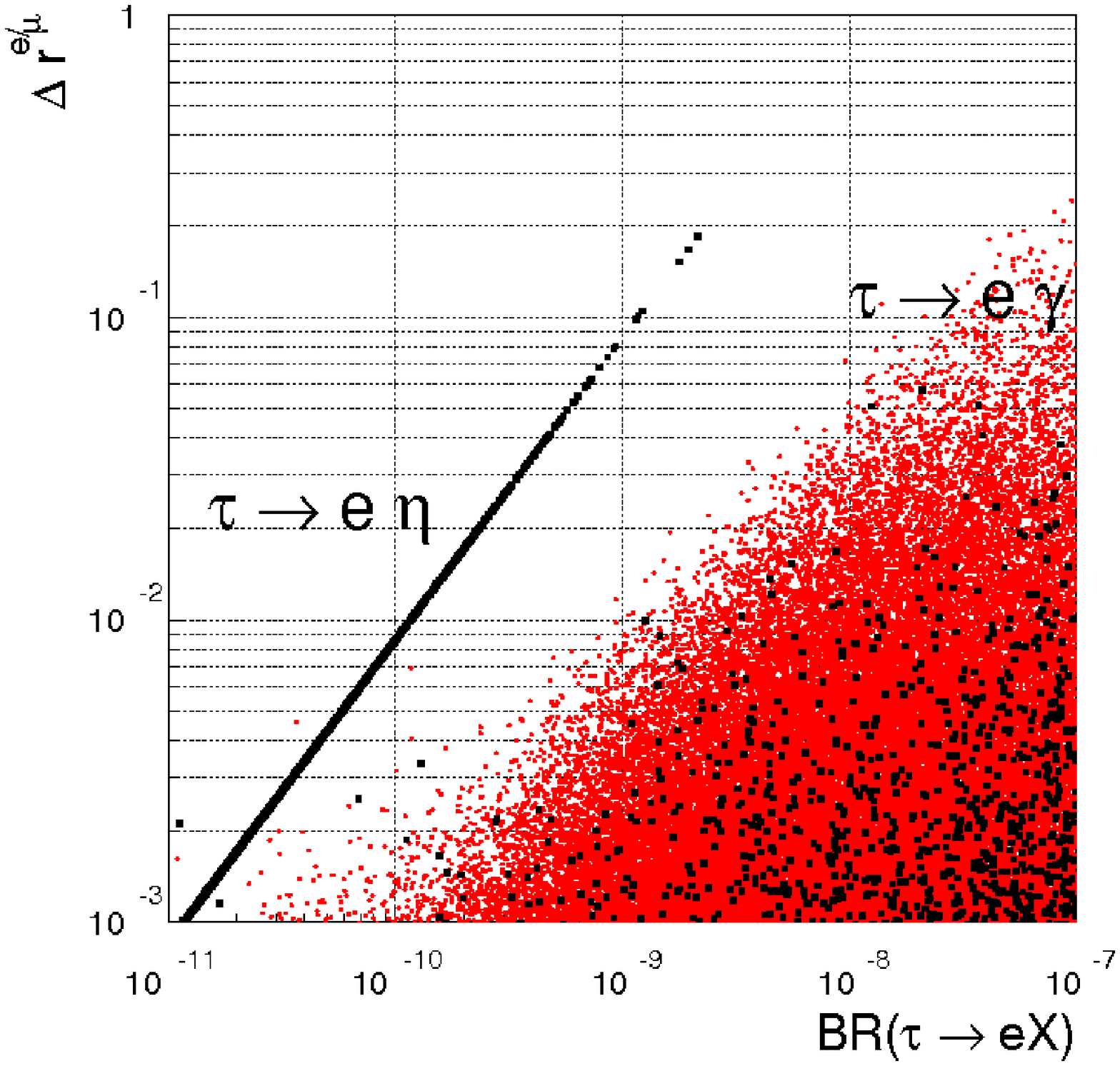}
\includegraphics[scale=0.38]{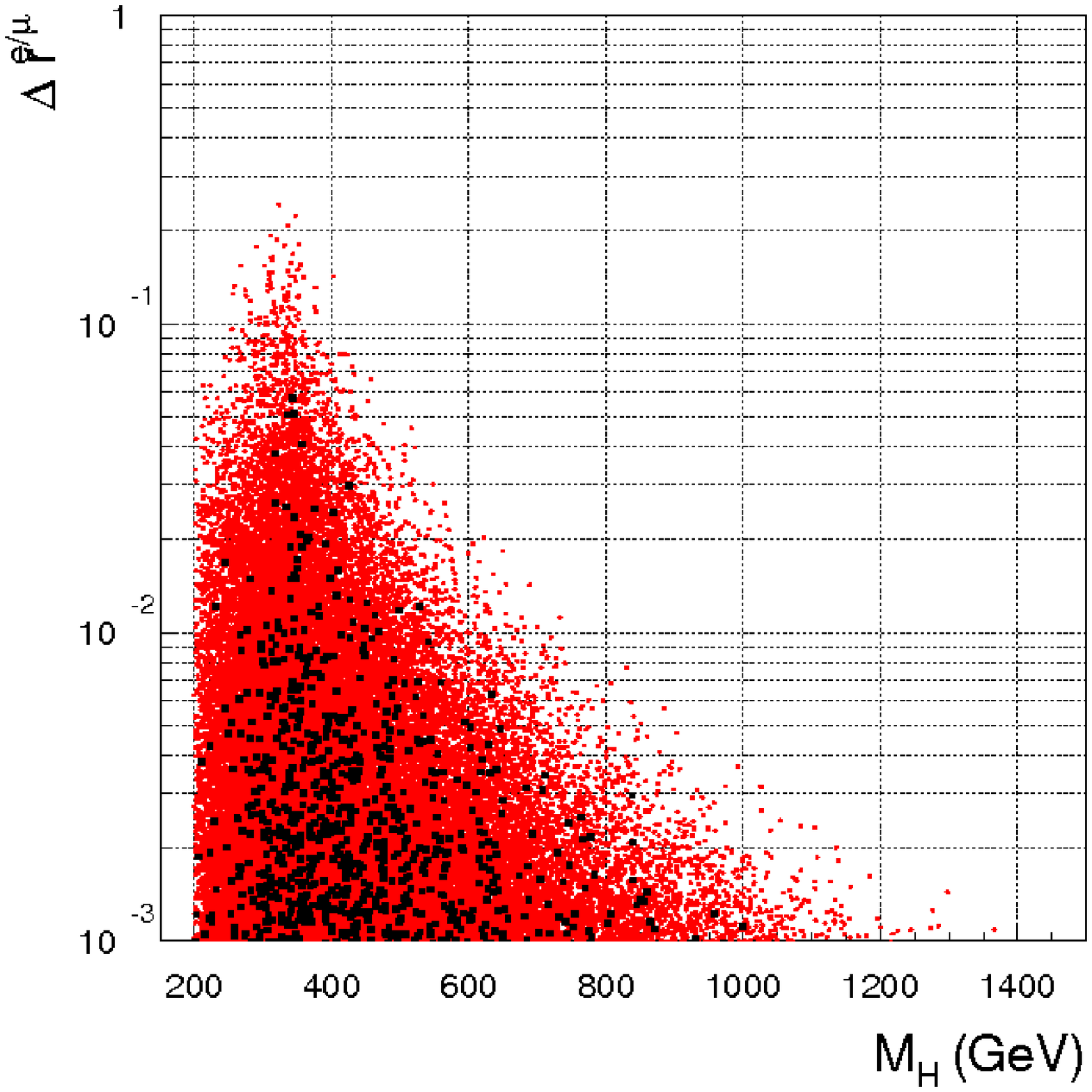}
\caption{\label{rklfv_R}
Left: $\Delta r_K^{e/\mu}$ as a function of $\BR(\tau\!\to\! e\gamma)$
and $\BR(\tau\!\to\!e\eta)$. Right: $\Delta r_K^{e/\mu}$ as a function
of $M_H$. Both plots have been obtained by means of a scan over the
following parameter space: $(m_{L,R}, m_{\tilde{Q}},
m_{\tilde{g}}, m_{\tilde{W}}, m_{\tilde{B}}, M_{H})<2.5\rm{TeV}$,
$\mu\!<\!5$\rm{TeV}, $|\delta_{RR}|\!<\!0.5$, $|\delta_{LL}|\!=\!0$ and
$\tan\beta<60$. Black dots refer to the points satisfying $1\times
10^{-9}<(g-2)_\mu <5\times 10^{-9}$.}
\end{figure}

Let us now discuss LFU breaking effects in $\Delta r_K^{e/\mu}$ as generated by LFV
contributions stemming, in particular, from \rm{RR}-type flavor violating sources only.
In Fig.~\ref{rklfv_R}, on the left, we report $\Delta r_K^{e/\mu}$ as a function of
$\BR(\tau\!\to\! e\gamma)$ and $\BR(\tau\!\to\!e\eta)$ while, on the right, we report
$\Delta r_K^{e/\mu}$ as a function of $M_H$. The plots have been obtained by means of
a scan over the following parameter space: $(m_{L,R}, m_{\tilde{Q}},
m_{\tilde{g}}, m_{\tilde{W}}, m_{\tilde{B}}, M_{H})<2.5\rm{TeV}$, $\mu\!<\!5$\rm{TeV}, $|\delta_{RR}|\!<\!0.5$, $|\delta_{LL}|\!=\!0$ and $\tan\beta<60$ and imposing all the
constraints discussed in Sec.\ref{constraints}.
Black dots refer to the points satisfying the $(g-2)_\mu$ anomaly at
the $95\%$ C.L., i.e. $1\times 10^{-9}<\Delta a_\mu<5\times 10^{-9}$.
Fig.~\ref{rklfv_R} clearly shows that there are quite a lot of points
in the interesting region where $0.001\!<\!\Delta r_K^{e/\mu}\!<\!0.01$
accounting for the $(g-2)_\mu$ anomaly and that are compatible with the 
experimental constraints of $\BR(\tau\!\to\!e\gamma)$ and $\BR(\tau\!\to\!e\eta)$.

As discussed in the previous section and as it is shown in Fig.~\ref{rklfv_R},
$\Delta r_K^{e/\mu}$ and $\BR(\tau\!\to\!e\eta)$ are closely related, at least 
in the limiting case where $\Delta_L =0$.
On the contrary, an analogue correlation between $\Delta r_K^{e/\mu}$ and 
$\BR(\tau\!\to\!e\gamma)$ is absent, due to their different dependence on
the SUSY mass spectrum.

\begin{figure}
\centering
\includegraphics[scale=0.38]{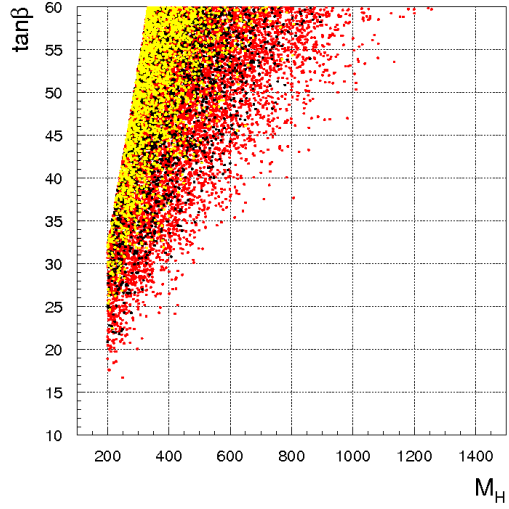}
\caption{\label{tgbmh_R}
Regions of the parameter space in the $\tan\beta-M_H$ plane
where $0.001\!<\!|\Delta r_K^{e/\mu}|\!<\! 0.003$ (red dots),
$0.003\!<\!|\Delta r_K^{e/\mu}|\!<\! 0.005$ (black dots) and
$|\Delta r_K^{e/\mu}|\!>\! 0.005$ (yellow dots)
as obtained by means of the same scan performed in Fig.~\ref{rklfv_R}.}
\end{figure}

\begin{figure}
\centering
\includegraphics[scale=0.38]{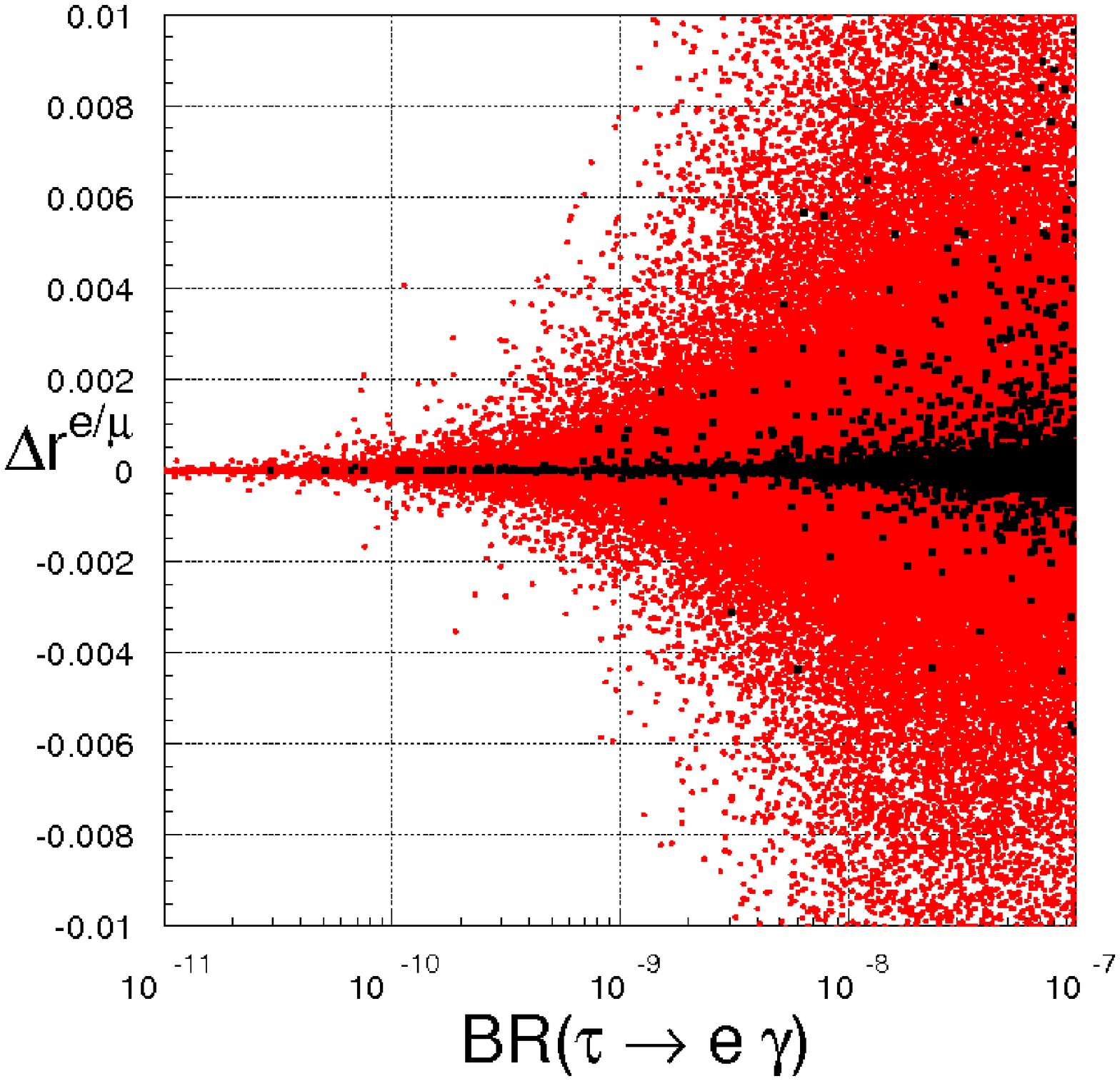}
\includegraphics[scale=0.38]{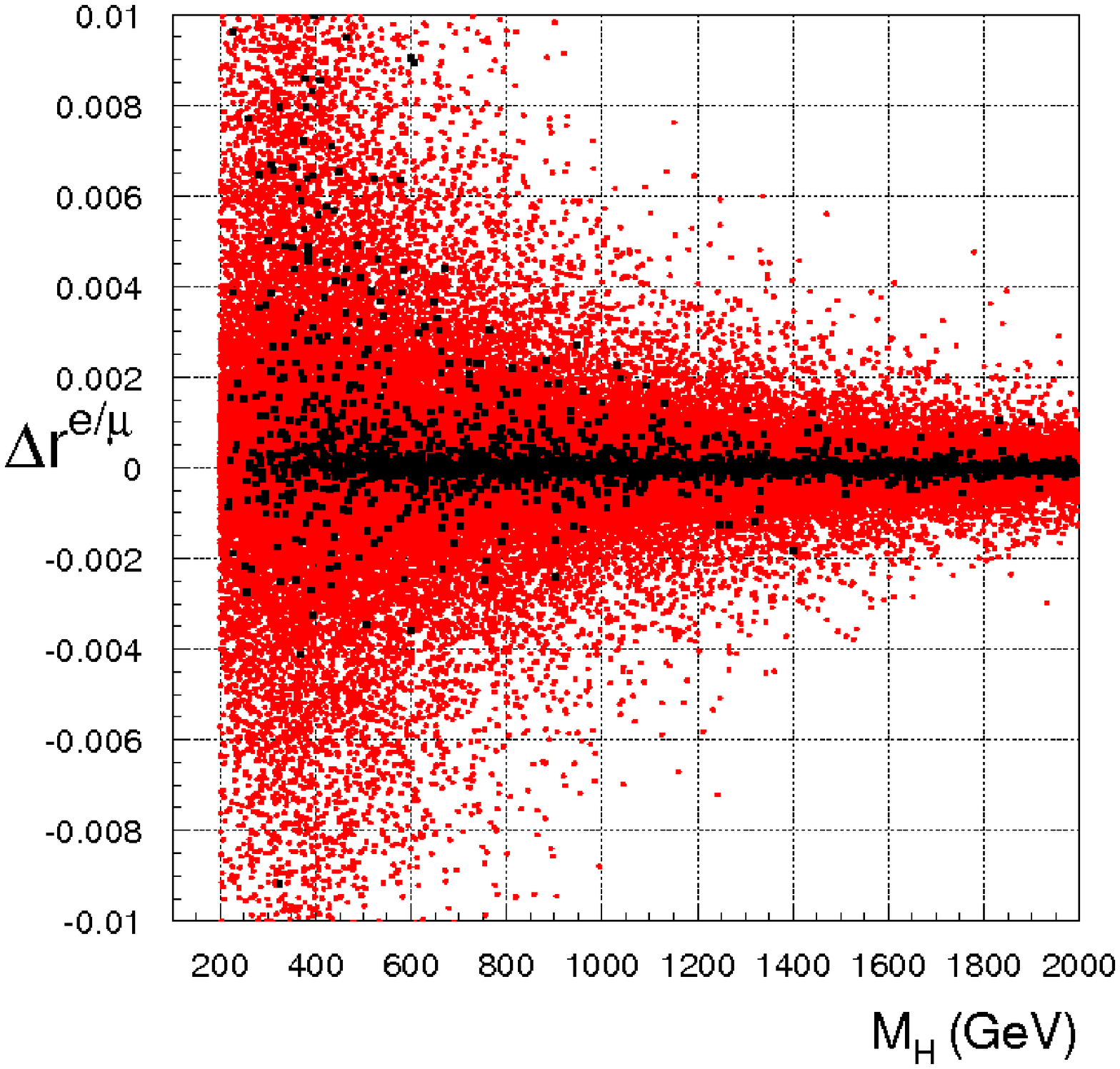}
\caption{\label{rklfv_LR}
Left: $\Delta r_K^{e/\mu}$ as a function of $\BR(\tau\!\to\! e\gamma)$
and $\BR(\tau\!\to\!e\eta)$. Right: $\Delta r_K^{e/\mu}$ as a function
of $M_H$. Both plots have been obtained by means of a scan over the
following parameter space: $(m_{L,R}, m_{\tilde{Q}}, m_{\tilde{g}}, m_{\tilde{W}},
m_{\tilde{B}}, M_{H})<2.5\rm{TeV}$, $\mu\!<\!5$\rm{TeV}, $|\delta_{RR,LL}|\!<\!0.5$
and $\tan\beta<60$. 
Black dots refer to the points satisfying $1\times 10^{-9}<(g-2)_\mu <5\times 10^{-9}$.}
\end{figure}

We also emphasize that experimentally visible effects in $\Delta r_K^{e/\mu}$
(at the $0.1\%$ level) can be reached up to charged Higgs masses at the TeV scale,
as shown in Fig.~\ref{rklfv_R} on the right. Moreover, we also stress that the
present experimental bounds on $\Delta r_K^{e/\mu}$ at the $\%$ level already set
constraints on the SUSY parameter space.
In Fig.~\ref{tgbmh_R}, we show the SUSY parameter space in the $\tan\beta-M_H$ 
plane probed by an experimental resolution on $\Delta r_K^{e/\mu}$ up to the 
$0.1\%$ level. In particular, red dots refer to the points satisfying
$0.001\!<\!|\Delta r_K^{e/\mu}|\!<\! 0.003$, black dots refer to the points where $0.003\!<\!|\Delta r_K^{e/\mu}|\!<\! 0.005$ and, finally, yellow dots are relative
to the points where $|\Delta r_K^{e/\mu}|\!>\! 0.005$.
As discussed in  Sec.~\ref{LFU_LFV}, it is also possible 
to generate LFU breaking effects in $\Delta r_K^{e/\mu}$ by means 
of a double source of LFV that, as a final result, preserve the lepton
flavor (see Eq.~\ref{LFUlfv}). This is the case when both LL and RR
flavor violating sources are simultaneously non vanishing. The major 
novelty arising from this last possibility is that now the new physics 
contributions can interfere with the SM ones; thus, we can get both
positive and negative values for $\Delta r_K^{e/\mu}$.
This is clearly shown by Fig.~\ref{rklfv_LR} that is the analog of
Fig.~\ref{rklfv_R} but in the presence of non vanishing $\delta_{LL}$
LFV terms. We see that $\Delta r_K^{e/\mu}$ can lie in the
experimentally interesting region while satisfying all the current
constraints. We observe that also in this case, the requirement of
large LFU breaking effects in $\Delta r_K^{e/\mu}$ at the level of
$0.001\!<\!|\Delta r_K^{e/\mu}|\!<\!0.01$, can be compatible with an
explanation for the $(g-2)_{\mu}$ anomaly while satisfying the constraints
from $\BR(\tau\!\to\! e\gamma)$ and $\BR(\tau\!\to\!e\eta)$.

\begin{figure}
\centering
\includegraphics[scale=0.38]{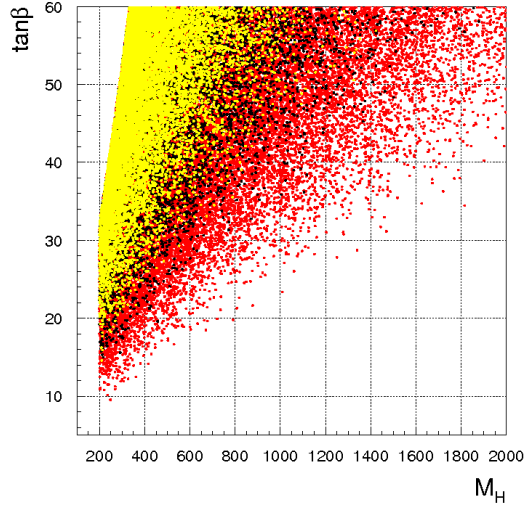}
\caption{\label{tgbmh_LR}
Regions of the parameter space in the $\tan\beta-M_H$ plane
where $0.001\!<\!|\Delta r_K^{e/\mu}|\!<\! 0.003$ (red dots),
$0.003\!<\!|\Delta r_K^{e/\mu}|\!<\! 0.005$ (black dots) and
$|\Delta r_K^{e/\mu}|\!>\! 0.005$ (yellow dots)
as obtained by means of the same scan performed in Fig.~\ref{rklfv_LR}.}
\end{figure}

Finally, Fig.~\ref{tgbmh_LR} shows the parameter space in the $\tan\beta-M_H$ 
plane probed by an experimental resolution on $\Delta r_K^{e/\mu}$ up to the 
$0.1\%$ level in analogy to Fig.~\ref{tgbmh_R}.
\section{2HDM framework}
\label{2HDM}
Theories with only one Higgs doublet, like the Standard Model (SM), do not
contain flavor violating interactions of the fermions with the Higgs bosons.
In particular, it is always possible to simultaneously diagonalize the
fermion mass matrices and the Higgs-fermion couplings.
In general, this is no longer true in models with several Higgs doublets.
In fact, up and down-type fermions can couple, at the same time, to more 
than a single scalar doublet and this naturally leads to FCNC effects at 
the tree level.
To suppress tree level FCNC in the theory so as not to be in conflict with
known experimental limits, an \textit{ad hoc} discrete symmetry is typically 
invoked. For instance, in the 2HDM, the up-type and the down-type quarks couple
either to the same Higgs doublet (this is known as the 2HDM-I)~\cite{Gunion:1989we},
or to different doublets (2HDM-II)~\cite{Gunion:1989we}.
On the other hand, in the most general case, the so-called 2HDM-III~\cite{2HDMIII},
no discrete symmetries are assumed and FCNC phenomena naturally appear.

The Lagrangian for the LFV Yukawa couplings of the 2HDM type III reads~\cite{2HDMIII}
\begin{equation}
-{\cal L}=\eta_{ij} {\bar \ell}_{iL} H_1 \ell_{jR} +\xi_{ij} {\bar \ell}_{iL}
H_2 \ell_{jR} + {\mathrm {h.c.}}
\end{equation}
where $H_{1,2}$ are the Higgs doublets defined by $H_1=(\phi_1^+\,\,\phi_1^0)$
and $H_2=(\phi_2^+\,\,\phi_2^0)$ while $\eta_{ij}$ and $\xi_{ij}$ are off-diagonal
$3\times 3$ matrices in the flavor space and $i$, $j$ are family indices.

Passing to the basis where the leptons are in mass eigenstates and expressing
the leptonic Lagrangian in terms of $\xi_{ij}$, one find that
\begin{eqnarray}
{\cal L} &=& -\frac{m_i}{v c_{\beta}}{\bar \ell}_i \ell_i
(s_{\alpha} h^0 - c_{\alpha} H^0)
+i \frac{m_i t_{\beta}}{v} {\bar \ell}_i  \gamma_5 \ell_i A^0
+\frac{m_i t_{\beta}}{\sqrt{2} v}
{\bar\nu}_i (1+\gamma_5)\ell_i H^+ \\
&-& \frac{m_i}{v c_{\beta}} {\bar\ell}_i \xi_{ij}\ell_j
\left[c_{\alpha-\beta} h^0+ s_{\alpha-\beta} H^0\right]
- \frac{im_i}{v c_{\beta}} {\bar\ell}_i\xi_{ij}\gamma_5\ell_j A^0 -
\frac{m_i}{\sqrt{2} v c_{\beta}}{\bar\nu}_i\xi_{ij}(1+\gamma_5)\ell_j H^+
+{\mathrm {h.c.}}
\label{2HDM_LFV_lagrangian}
\end{eqnarray}
where $v=\sqrt{v_1^2+v_2^2}=(\sqrt{2}G_F)^{-1/2}=246$~GeV. Note that the Lagrangian (\ref{2HDM_LFV_lagrangian}), expressed in terms of the $\xi_{ij}$ matrices, has the 
same lepton flavor conserving interactions of the to 2HDM-II.
To be consistent with experimental data on FCNC processes, Cheng and Sher (CS),
inspired by the hierarchy in the fermion masses, have proposed the following
famous ansatz for the couplings $\xi_{ij}$~\cite{chengsher}:
\begin{equation}
\label{eq:coupl}
\xi_{ij} = \lambda_{ij}\sqrt{\frac{m_j}{m_i}},
\end{equation}
where the residual arbitrariness of flavor changing couplings is expressed by the
parameters $\lambda_{ij}$ which is constrained by experimental bounds on LFV processes.

By making use of the effective Lagrangion of Eq.~\ref{2HDM_LFV_lagrangian}, it is
straightforward to compute the expression for the quantity $\Delta r^{e/\mu}_{K}$.
It turns out that
\begin{eqnarray}
\label{LFUlfv2HDM}
\Delta r^{e/\mu}_{K}
&\simeq&
\bigg(\frac{m^{4}_{K}}{M^{4}_{H}}\bigg)
\bigg(\frac{m_{\tau}}{m_{e}}\bigg)^2
|\xi_{31}|^2t_{\beta}^{4}\\
&\simeq&
10^{-2}\times\bigg(\frac{500\rm{GeV}}{M_{H}}\bigg)^{4}
\bigg(\frac{t_{\beta}}{40}\bigg)^{4}
|\lambda_{31}|^2\,,
\label{LFUlfv2HDM_CS}
\end{eqnarray}
where in Eq.~\ref{LFUlfv2HDM_CS} we made use of the the CS ansatz;
Eq.~\ref{LFUlfv2HDM_CS} clearly shows that a 2HDM of type III, with
the addition of the CS ansatz, can naturally predict a LFU breaking
in the $K\to\ell\nu$ systems at a visible level for natural values of
$M_{H}$ and $t_{\beta}$.
We also remind that the $\lambda_{ij}$ parameters should be typically of order
one~\cite{chengsher}. However, once we assume the CS ansatz, we are naturally
lead with stringent correlations among all the LFV transitions, as for example
$\mu\to e\gamma$, $\mu+N\to e+N$, $\tau\to e\gamma$, $\tau\to\mu\gamma$ and so on.

We have explicitly checked that the precision test provided by LFU breaking
effects in $\Delta r^{e/\mu}_{K}$ represents the most powerful probe of the
CS ansatz, at least in the decoupling regime (where $M_H \simeq M_A$ and where
the lightest higgs boson doesn't have LFV couplings with fermions)
and assuming non vanishing LFV interactions only for $3j$ transitions.

Irrespective of the specific model one can assume, we wish to emphasize  that
LFU breaking effects in $\Delta r^{e/\mu}_{K}$ represent the best probe for
$31$ LFV transitions in a generic 2HDM with tree level LFV couplings
(see Eq.~\ref{2HDM_LFV_lagrangian}).
To see this point explicitly, it is natural to compare the New Physics sensitivity
in LFU violation to that achievable in  $\tau\rightarrow e\eta$. Indeed,
this latter decay channel represents the most sensitive channel to New Physics
among all the rare $\tau$ decays (this is strictly true in the decoupling limit).
It turns out that
\beq
Br(\tau\rightarrow e\eta)
\simeq
10^{-8}\times\Delta r^{e-\mu}_{\!K}
\label{rktmueta_2HDM}
\eeq
showing that, within a 2HDM-III framework, $\Delta r^{e-\mu}_{\!K}$
sets tight constraints on the observation of $\tau\rightarrow e\eta$ at the
level of $Br(\tau\rightarrow e\eta)\leq 10^{-10}$.

\section{Conclusions}
\label{conclusion}
High precision electroweak tests, such as deviations from the Standard Model
expectations of the Lepton Universality breaking, represent a powerful tool 
to test the Standard Model and, hence, to constrain or obtain indirect hints
of new physics beyond it. Kaon and pion physics are obvious grounds where to 
perform such tests, for instance in the $K\rightarrow \ell\nu_{\ell}$ decays,
where $l= e$ or $\mu$.

In this paper, we have analyzed the domain of $\Delta r_K^{e/\mu}$ between
$10^{-3} <\Delta r_K^{e/\mu}< 10^{-2}$. An evidence of LFU violation at the
level of $\Delta r_K^{e/\mu}$ larger than $5\times 10^{-3}$ unambiguously
points towards the presence of LFV sources. On the other hand, if our increased
experimental sensitivity allows us to observe an LFU violation with values of
$\Delta r_K^{e/\mu}$ smaller than $5\times 10^{-3}$, then both the flavor
conserving and the flavor changing sources of LFU violation can be at play.
In any case, the observation of a non-vanishing $\Delta r_K^{e/\mu}$ in the 
range $10^{-3} <\Delta r_K^{e/\mu}< 5\times 10^{-3}$ would severely limit 
values in the $M_H-\tan\beta$ plane. If a signal exists at a such a level,
the LHC results become the crucial tool to disentangle flavor conserving and
flavor changing sources of LFU violation.

Interestingly enough, a process that in itself does not need lepton flavor violation
to occur, i.e. the violation of $\mu-e$ non-universality in $K_{\ell2}$, proves to 
be quite effective in constraining not only relevant regions of SUSY models where 
lepton flavor is conserved, but even those where specific lepton flavor violating 
contributions arise. Indeed, a comparison with analogous bounds coming from $\tau$
Lepton Flavor Violation decays shows the relevance of the measurement of $R_{K}$ to
probe Lepton Flavor Violation in SUSY.\\

\textit{Acknowledgments:}
This work has been supported in part by the Cluster of Excellence ``Origin and Structure of the
Universe'' and by the German Bundesministerium f{\"u}r Bildung und Forschung under contract 05HT6WOA.
One of us (A. M.) acknowledges Research Grants from the Istituto Nazionale di Fisica Nucleare
(INFN), project FA51 and from the Ministero dell' Istruzione, dell'Universita' e della Ricerca
(MIUR) and Univ. of Padova, astroparticle research project $2006023491-001$.
This research was supported in part by the European Community Research Training Networks under
contracts MRTN-CT-2004-503369 and MRTN-CT-2006-035505.
One of us (P. P.) acknowledges Andrzej J.~Buras for reading the paper and for useful discussions.

\end{document}